\documentclass[%
 reprint,
 superscriptaddress,
 altaffilletter
 amsmath,amssymb,
 aps,
]{revtex4-2}

\usepackage[utf8]{inputenc}
\usepackage[T1]{fontenc}
\usepackage[margin=1in]{geometry}
\usepackage{booktabs}
\usepackage{amsmath}
\usepackage{xcolor}
\usepackage[normalem]{ulem}

\usepackage{graphicx}
\usepackage{bm}
\usepackage{etoolbox}

\begin{document}

\preprint{APS/123-QED}

\title{Framework for an AI-Ready ARPES Community Database}

\author{Zhi-Xun Shen}
    \email{Corresponding author: zxshen@stanford.edu} 
    \affiliation{Physics Department, Stanford University, Stanford, California 94305, USA}
    \affiliation{Applied Physics Department, Stanford University, Stanford, California 94305, USA}
    \affiliation{Stanford Institute for Materials and Energy Sciences, SLAC National Accelerator Laboratory, Menlo Park, California 94025, USA}
    \affiliation{Geballe Laboratory for Advanced Materials (GLAM), Stanford University, Stanford, CA 94305, USA}
    \affiliation{Stanford Laboratory for Intelligent Matter, Stanford University, Stanford, CA 94305, USA}
\author{Jin Gene Wong}
    \affiliation{Physics Department, Stanford University, Stanford, California 94305, USA}
    \affiliation{Stanford Institute for Materials and Energy Sciences, SLAC National Accelerator Laboratory, Menlo Park, California 94025, USA}
    \affiliation{Geballe Laboratory for Advanced Materials (GLAM), Stanford University, Stanford, CA 94305, USA}
    \affiliation{Stanford Laboratory for Intelligent Matter, Stanford University, Stanford, CA 94305, USA}
\author{Karen Lei}
    \affiliation{Physics Department, Stanford University, Stanford, California 94305, USA}
    \affiliation{Stanford Institute for Materials and Energy Sciences, SLAC National Accelerator Laboratory, Menlo Park, California 94025, USA}
    \affiliation{Geballe Laboratory for Advanced Materials (GLAM), Stanford University, Stanford, CA 94305, USA}
    \affiliation{Stanford Laboratory for Intelligent Matter, Stanford University, Stanford, CA 94305, USA}
\author{León Garcia}
    \affiliation{Physics Department, Stanford University, Stanford, California 94305, USA}
    \affiliation{Stanford Laboratory for Intelligent Matter, Stanford University, Stanford, CA 94305, USA}
\author{Felix Baumberger}
    \affiliation{Department of Quantum Matter Physics, University of Geneva, 24 Quai Ernest-Ansermet, 1211 Geneva 4, Switzerland}
    \affiliation{Swiss Light Source, Paul Scherrer Institut, CH-5232 Villigen PSI, Switzerland}
\author{Sergey Borisenko}
    \affiliation{Leibniz Institute for Solid State and Materials Research, IFW Dresden, 01069 Dresden, Germany}
\author{Aaron Bostwick}
    \affiliation{Advanced Light Source, Lawrence Berkeley National Laboratory, Berkeley, California 94720, USA}
\author{Ting Cao}
    \affiliation{Department of Materials Science and Engineering, University of Washington, Seattle, Washington 98195, USA}
\author{Tanny Chavez}
    \affiliation{Advanced Light Source, Lawrence Berkeley National Laboratory, Berkeley, California 94720, USA}
\author{Yulin Chen}
    \affiliation{Department of Physics, Clarendon Laboratory, University of Oxford, Parks Road, Oxford OX1 3PU, UK}
\author{Monika Choudhary}
    \affiliation{Advanced Light Source, Lawrence Berkeley National Laboratory, Berkeley, California 94720, USA}
\author{Andrea Damascelli}
    \affiliation{Department of Physics \& Astronomy, University of British Columbia, Vancouver, British Columbia V6T 1Z1, Canada}
    \affiliation{Quantum Matter Institute, University of British Columbia, Vancouver, British Columbia, V6T 1Z4, Canada}
\author{Dan Dessau}
    \affiliation{Department of Physics, University of Colorado Boulder, Boulder, Colorado 80309, USA}
\author{Thomas P. Devereaux}
    \affiliation{Stanford Institute for Materials and Energy Sciences, SLAC National Accelerator Laboratory, Menlo Park, California 94025, USA}
    \affiliation{Department of Materials Science and Engineering, Stanford University, Stanford, California 94305, USA}
    \affiliation{Geballe Laboratory for Advanced Materials, Stanford University, Stanford, California 94305, USA}
\author{Sandy A. Ekahana}
    \affiliation{Department of Physics, Carnegie Mellon University, Pittsburgh, Pennsylvania 15213, USA}
    \affiliation{Advanced Light Source, Lawrence Berkeley National Laboratory, Berkeley, California 94720, USA}
\author{Alexei Fedorov}
    \affiliation{Advanced Light Source, Lawrence Berkeley National Laboratory, Berkeley, California 94720, USA}
\author{Atsushi Fujimori}
    \affiliation{Department of Physics, The University of Tokyo, 7-3-1 Hongo, Bunkyo-ku, Tokyo 113-0033, Japan}
    \affiliation{National Synchrotron Radiation Research Center, Hsinchu 300092, Taiwan}
    \affiliation{Center for Quantum Technology, and Department of Physics, National Tsing Hua University, Hsinchu 300044, Taiwan}
\author{Yucheng Guo}
    \affiliation{Department of Physics and Astronomy, Rice University, Houston, Texas 77251, USA}
\author{M. Zahid Hasan}
    \affiliation{Laboratory for Topological Quantum Matter and Advanced Spectroscopy, Department of Physics, Princeton University, Princeton, New Jersey 08544, USA}
    \affiliation{Princeton Materials Institute, Princeton University, Princeton, New Jersey 08544, USA}
\author{Makoto Hashimoto}
    \affiliation{Stanford Synchrotron Radiation Lightsource, SLAC National Accelerator Laboratory, Menlo Park, California 94025, USA}
\author{Yu He}
    \affiliation{Department of Applied Physics, Yale University, New Haven, Connecticut 06511, USA}
\author{Philip Hofmann}
    \affiliation{Department of Physics and Astronomy, Aarhus University, 8000 Aarhus C, Denmark}
\author{Zahid Hussain}
    \affiliation{Materials Sciences Division, Lawrence Berkeley National Laboratory, Berkeley, California 94720, USA}
\author{Harold Y. Hwang}
    \affiliation{Stanford Institute for Materials and Energy Sciences, SLAC National Accelerator Laboratory, Menlo Park, California 94025, USA}
    \affiliation{Applied Physics Department, Stanford University, Stanford, California 94305, USA} 
\author{Zhurun Ji}
    \affiliation{Department of Physics, Massachusetts Institute of Technology, Cambridge, Massachusetts 02139, USA}
\author{Chunjing Jia}
    \affiliation{Department of Physics, University of Florida, Gainesville, Florida 32611, USA}
\author{Timur K. Kim}
    \affiliation{Diamond Light Source Ltd, Harwell Science and Innovation Campus, Didcot OX11 0DE, UK}
\author{Phil D. C. King}
    \affiliation{School of Physics and Astronomy, University of St Andrews, St Andrews KY16 9SS, UK}
\author{Takeshi Kondo}
    \affiliation{Institute for Solid State Physics, The University of Tokyo, Kashiwa, Chiba 277-8581, Japan}
\author{Alessandra Lanzara}
    \affiliation{Materials Sciences Division, Lawrence Berkeley National Laboratory, Berkeley, California 94720, USA}
    \affiliation{Department of Physics, University of California, Berkeley, Berkeley, California 94720, USA}
\author{Chun Lin}
    \affiliation{Stanford Synchrotron Radiation Lightsource, SLAC National Accelerator Laboratory, Menlo Park, California 94025, USA}
\author{Donghui Lu}
    \affiliation{Stanford Synchrotron Radiation Lightsource, SLAC National Accelerator Laboratory, Menlo Park, California 94025, USA}
\author{Matteo Michiardi}
    \affiliation{Quantum Matter Institute, University of British Columbia, Vancouver, British Columbia, V6T 1Z4, Canada}
\author{Sung-Kwan Mo}
    \affiliation{Advanced Light Source, Lawrence Berkeley National Laboratory, Berkeley, California 94720, USA}
\author{Brian Moritz}
    \affiliation{Stanford Institute for Materials and Energy Sciences, SLAC National Accelerator Laboratory, Menlo Park, California 94025, USA}
\author{Laurenz Rettig}
    \affiliation{Fritz Haber Institute of the Max Planck Society, Faradayweg 4-6, 14195 Berlin, Germany}
\author{Eli Rotenberg}
    \affiliation{Advanced Light Source, Lawrence Berkeley National Laboratory, Berkeley, California 94720, USA}
\author{Takafumi Sato}
    \affiliation{Department of Physics, Graduate School of Science, Tohoku University, Sendai 980-8578, Japan}
    \affiliation{International Center for Synchrotron Radiation Innovation Smart (SRIS), Tohoku University, Sendai 980-8577, Japan}
    \affiliation{Mathematical Science Center for Co-creative Society (MathCCS), Tohoku University, Sendai 980-8578, Japan}
    \affiliation{Advanced Institute for Materials Research (WPI-AIMR), Tohoku University, Sendai 980-8577, Japan}
    \affiliation{Center for Science and Innovation in Spintronics (CSIS), Tohoku University, Sendai 980-8577, Japan}
\author{Kyle M. Shen}
    \affiliation{Laboratory of Atomic and Solid State Physics, Department of Physics, Cornell University, Ithaca, New York 14853, USA}
    \affiliation{Kavli Institute at Cornell for Nanoscale Science, Ithaca, New York 14853, USA}
\author{Jonathan A. Sobota}
    \affiliation{Stanford Institute for Materials and Energy Sciences, SLAC National Accelerator Laboratory, Menlo Park, California 94025, USA}
\author{Inna Vishik}
    \affiliation{Department of Physics and Astronomy, University of California Davis, Davis, California 95616, USA}
    \affiliation{Materials Sciences Division, Lawrence Berkeley National Laboratory, Berkeley, California 94720, USA}
\author{Shuolong Yang}
    \affiliation{Pritzker School of Molecular Engineering, University of Chicago, Chicago, Illinois 60637, USA}
\author{Ming Yi}
    \affiliation{Department of Physics and Astronomy, Rice University, Houston, Texas 77251, USA}
\author{Timothy Zaklama}
    \affiliation{Department of Physics, Massachusetts Institute of Technology, Cambridge, Massachusetts 02139, USA}

\begin{abstract}
Condensed matter stands to gain profoundly from widespread adoption of artificial intelligence (AI). However, its current means of data organization impedes this endeavor. As a premier measure of electronic structure underlying material properties, Angle-Resolved Photoemission Spectroscopy (ARPES) has accumulated over four decades of spectra, yet the community's publication-centered workflow preserves only a highly selected fraction of its experimental record. A shared data ecosystem could transform these fragmented measurements and their contexts into a persistent, collective scientific memory, creating immediate value for reproducibility and reuse while laying the foundation for AI-enabled discovery. In this Perspective, we report outcomes of a recent community workshop and propose a practical path toward such an ecosystem. Its benefits unfold along three timescales. In the near term, standardized, searchable, and citable datasets strengthen analysis pipelines independently of AI. In the intermediate future, community-validated tools operating across laboratories could enable broader reuse and increasingly closed-loop automated experiments. In the long term, a diverse corpus of measured electronic spectra could support foundation models for quantum matter. Looking forward, a diverse corpus of measured electronic spectra could support foundation models for quantum matter and new discoveries. We discuss lessons from existing scientific databases, the current state of AI in ARPES, as well as areas of community convergence on data organization, access, credit, governance, and a staged pilot program.
\end{abstract}

\maketitle

Artificial intelligence (AI) is transforming how scientific inquiry is conducted across an expanding range of disciplines. Data, and the context needed to interpret them, have played a decisive role in this transformation. Theoretical research benefits from extensive databases of calculations; structural sciences have mature data repositories amassed over decades, such as the Protein Data Bank for biology. In contrast, experimental condensed matter physics largely lacks standardized data repositories, instead responding to recent data availability requirements in an ad hoc manner. Organizing community-wide data no longer just serves to ensure reproducibility, it also becomes the key step to enable AI-native science in the future. The community needs to organize their data now — not with AI initially in mind, but simply to make today’s science more reproducible and reusable.

Many of the most profound questions in condensed matter can only be answered by reasoning over broad classes of materials, an undertaking for which today's single-system approach to data collection, analysis, and publication proves inadequate. This calls for large community repositories of experimental data to be assembled, in a manner that both fortifies the current scientific workflow and sets the ground for deeper AI adoption. Angle-Resolved Photoemission Spectroscopy (ARPES), with its information-rich measurements and close connection to the microscopic “ground truth” of effective lattice models, provides a timely test case for changing the modus operandi. This offers the opportunity to transform decades of fragmented experimental experience into a persistent, computationally accessible collective memory of electronic structure. A recent workshop at Stanford University brought together members of the ARPES community to explore this opportunity \cite{aies2026}. This Perspective draws on discussions at that workshop to present a viable road map for the community, and as a request for the rest of the community to join our effort. We examine why ARPES data are scientifically distinctive, what is lost under current data practices, what new forms of science the ecosystem could enable, and what principles should guide its construction. We note that not all matters were resolved at the conference — consultation with the ARPES community, guided by a steering committee (Supplementary Section~S8) elected by conference participants, and the creation of pilot databases to catch and rectify issues are all necessary prerequisites before the ecosystem may be scaled up. This paper has three primary intents: (1) to convince the ARPES community of the value of this endeavor not just by laying out a vision in this paper, (2) to invite all interested parties to our follow-up workshops on future steps, and (3) highlight concrete demonstrations of its potential through pilot projects deployed in the immediate future.

\section{Why electronic structure matters, and the ARPES opportunity}

Electronic structure is the \textit{microscopic} physical foundation that underpins many crucial material properties. By describing the behavior of quasi-particles (which emerge from electrons coupling to spin, lattice, and orbital degrees of freedom) it determines important \textit{macroscopic} properties such as the development of collective order, bulk transport properties (including conductivity, magnetism, and optical response), as well as catalytic properties. Electronic and crystal structures are strongly interdependent and may be jointly viewed as the ``genome'' of a material; while not a description of a system in its entirety, they are a compact microscopic basis from which physical properties could be understood.

\begin{figure*}[htbp]
\centering
\includegraphics[width=0.85\textwidth]{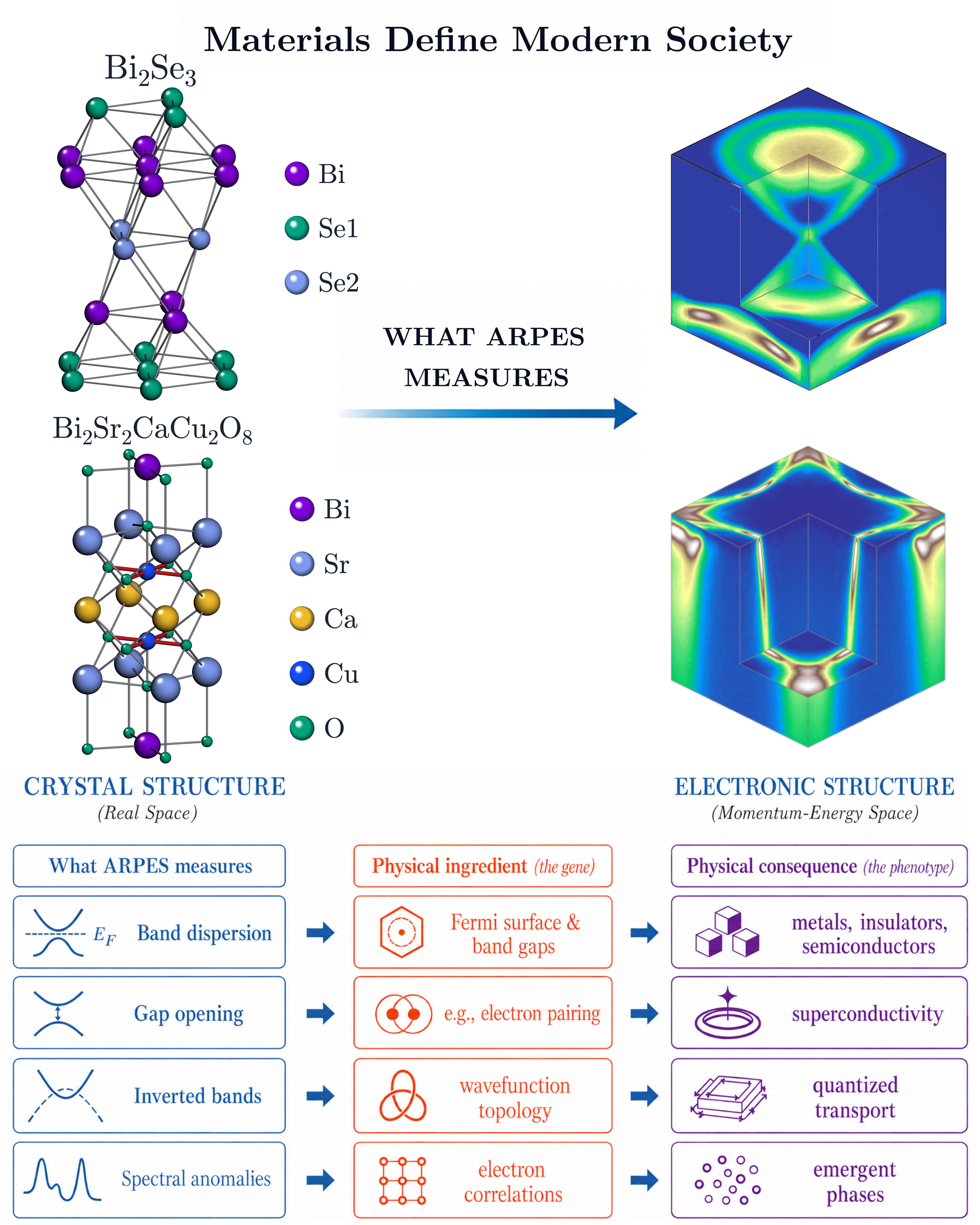}
\caption{Electronic structure is the genome of quantum materials, and ARPES reads it. Top: for a given crystal structure in real space (left), ARPES measures the momentum-resolved electronic structure — the photoemission intensity $\mathcal{I}(k_x, k_y, E)$ in momentum–energy space (right). Bottom: each class of spectral observable (left column) exposes a physical ingredient of the material, its ``gene'' (center), which helps determines a macroscopic consequence, the ``phenotype'' (right): band dispersions locate the Fermi surfaces and band gaps that separate metals, insulators, and semiconductors; gap openings signal the electron pairing behind superconductivity; bands inverts encode the wave function topology in spin-orbit coupled systems behind quantized transport; and spectral anomalies reveal the electron correlations behind emergent phases. ARPES intensity volume adapted from data in \cite{lee2007} with permission.}
\label{fig:overview}
\end{figure*}

ARPES provides direct access to electronic structure by measuring the energy and momenta of photoemitted electrons (Fig.~\ref{fig:overview}) \cite{damascelli2003,sobota2021}. These electrons are collected by analyzers, which generate spectra from the measurements, giving insight into microscopic ingredients that thus determine macroscopic properties. For conventional equilibrium measurements, ARPES primarily probes the occupied one-electron spectral function, modulated by photoemission matrix elements and experimental resolution. While pump-probe approaches and inverse photoemission extend access above the Fermi level, conventional ARPES comprises the vast majority of experiments performed, and is thus the focus of this Perspective. 

Because ARPES resolves band dispersions and many-body effects, it has played a central role in the study of major quantum materials such as cuprate and iron-based superconductors, topological materials, and two-dimensional systems \citep{damascelli2003,sobota2021,basov2014rmp,hasan2010rmp}. Coming a long way since its emergence in the 1970s, it has matured into a broadly quantitative, high-resolution probe of many-body electronic structure \cite{smith1974, eberhardt1980}. As shown in Fig.~\ref{fig:landscape}, bursts of activity followed high-resolution studies of the cuprates in the early 1990s, exemplified by the confirmation of $d$-wave pairing \cite{shen1993}, and the spread of ARPES from the late 2000s to graphene \cite{ohta2006,zhou2007}, topological \cite{hsieh2008,xia2009,chen2009}, two-dimensional \cite{zhang2014,zhang2016,tang2017}, and other quantum materials \cite{xu2015,he2013,ding2008}, powered by development from hemispherical analyzers \cite{roy1990,martensson1994} to new two-dimensional imaging \cite{borisenko2022}, and multidimensional data collection schemes \cite{schonhense2020}. New modalities including momentum microscopy \cite{kromker2008,schonhense2020,medjanik2017}, time \cite{boschini2024}, spin \cite{hsieh2008,hsieh2009}, and spatial \cite{avila2013,rotenberg2014} resolution, and magnetic fields \cite{ryu2023, huang2023} continue to expand both the scientific scope of the technique and the volume and dimensionality of the data it produces.

\section{From complicated and fragmented data — to collective scientific memory}

\subsection{ARPES data are extremely rich, but they come with inherent complexity}

Although ARPES is an extremely powerful experimental technique, data interpretation is complicated. It measures electronic structure, but its recorded intensity is not the theoretical spectral function (Table~\ref{tab:arpes}) — it depends strongly on \textit{experimental context}, including experimental geometry, instrumental resolution and detector response, coordinate transformations, data-processing choices, and (perhaps most crucially) exceptional sensitivity to sample surface conditions at the time of measurement. Like scattering techniques such as X-ray diffraction, ARPES samples only \textit{slices of} a high-dimensional momentum space, whose exact locations in that space are revealed in post-processing. For these reasons, measurements that should capture identical theoretical spectra can look substantially different across beamlines, samples, locations on a sample, or times after cleavage. The central data challenge is therefore to \textit{preserve and standardize the experimental context} required for successful interpretation.

\begin{table*}
\centering
\footnotesize
\caption{Complications in ARPES and their ramifications.}
\label{tab:arpes}
\renewcommand{\arraystretch}{1.25}
\begin{tabular}{p{4.8cm} p{11.2cm}}
\toprule
\textbf{ARPES context/development} & \textbf{What should be preserved or accommodated}  \\
\toprule
Photoemission matrix elements & \raggedright Photon energy, polarization, experimental geometry, other parameters controlling measured intensity \tabularnewline \midrule
Instrument response & \raggedright Energy/angular resolution, detector characteristics, calibration, and analyzer settings  \tabularnewline \midrule
High-dimensional sampling & \raggedright Raw measured coordinates and information describing how the underlying multidimensional space was sampled \tabularnewline \midrule
Sample environment & \raggedright For example, applied fields, strain, currents, optical excitation, correlating ARPES to transport measurements; should preserve information on instruments used for setting and measuring these conditions, and sample responses \tabularnewline \midrule
Coordinate conversion & \raggedright Emission angles, manipulator conventions, photon geometry, analyzer configuration, and transformations into momentum space  \tabularnewline \midrule
Surface sensitivity & \raggedright Sample identity, cleavage history, measurement location, temperature/time history, aging, and spectral-quality evolution, including XPS reference spectra where available  \tabularnewline \midrule
Data processing & \raggedright Processing steps and parameters, while retaining access to less-processed or raw representations  \tabularnewline \midrule
Increasing dimensionality & \raggedright Evolution from 1D spectra $\rightarrow$ 2D energy–momentum maps $\rightarrow$ full momentum/TOF mapping $\rightarrow$ spatial (micro- and nano-ARPES), spin, and time-resolved datasets  \tabularnewline \midrule
Cross-instrument variability & \raggedright Sufficient provenance to compare measurements across beamlines, laboratories, instruments, and experimental sessions  \tabularnewline \midrule
Metadata standards & \raggedright A common community-defined description of signal, coordinates, experimental conditions, and provenance  \tabularnewline \midrule
Open-data practice & \raggedright Harmonization of currently piecemeal publication-linked repositories into a searchable, reusable community resource  \tabularnewline
\bottomrule
\end{tabular}
\end{table*}

\subsection{The current ARPES workflow is suboptimal}

\begin{figure*}[htbp]
\centering
\includegraphics[width=0.8\textwidth]{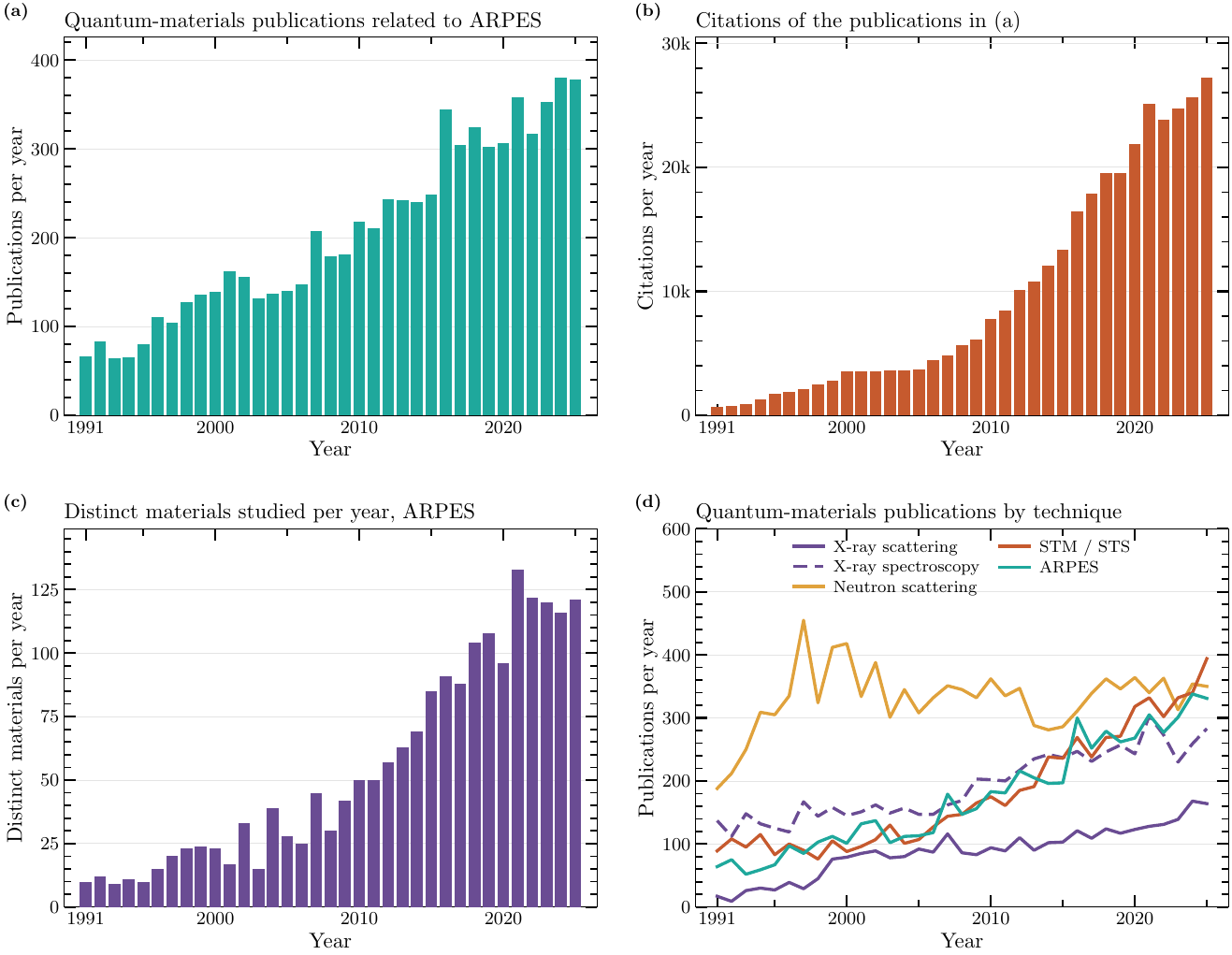}

\caption{Growth of ARPES activity, impact, and material coverage in quantum materials, from Web of
Science (current as of September 2026 with totals cover 1991-2025).
(a)~Publications per year matching an All Databases search for
``angle-resolved photoemission'' and its spelling and acronym variants
(ARPES, angle resolved photo emission, etc.), restricted by a keyword filter to quantum-materials topics (correlated electrons and superconductivity, topology, quantum magnetism, and moir\'e and kagome lattices) and excluding preprints, Research
Commons, datasets, grants, and patents; 7,188
records in total.
(b)~Citations received in each calendar year by the publications in (a),
341,862 in total. Citations are binned by the year in which they accrue, so recent years are not suppressed by citation lag.
(c)~Distinct materials named in the titles of the publications in (a). Materials are identified by parent compound and distinct stoichiometries count separately
($\mathrm{Bi}_2\mathrm{Sr}_2\mathrm{CuO}_6$ vs.\
$\mathrm{Bi}_2\mathrm{Sr}_2\mathrm{CaCu}_2\mathrm{O}_8$), doping and strain
collapse onto the parent ($\mathrm{La}_{2-x}\mathrm{Sr}_x\mathrm{CuO}_4$
counts as $\mathrm{La}_2\mathrm{CuO}_4$ at every $x$), solid solutions
resolve onto their independently studied endmembers ($\mathrm{FeSe}_{1-x}\mathrm{Te}_x$ counts as FeSe and FeTe), and a film on a substrate is distinct from the bulk ($\mathrm{FeSe/SrTiO}_3$ vs.\ FeSe). Since only titles are parsed, these counts are approximate; 726 distinct materials appear in total.
(d)~Publications per year for five experimental probes of quantum materials
under equivalent searches, each for the technique's name as an exact phrase
without acronyms and restricted by the same filter, so the ARPES curve holds 85\% of the publications in (a). X-ray spectroscopy covers X-ray absorption, emission and dichroism; resonant inelastic X-ray scattering is counted under X-ray scattering, and X-ray photoelectron spectroscopy is not included.
The X-ray and neutron search strings match
families of related techniques whereas the ARPES string matches a single
technique, so cross technique ratios are indicative rather than exact.
All panels begin in 1991, the first year the Web of Science Core Collection
indexes abstracts and author keywords; a title-only control confirms that the large apparent 1990-91 jump in its topic-search counts is an indexing artifact, and earlier counts, which mostly reflect title wording, are omitted.}
\label{fig:landscape}
\end{figure*}

\begin{figure*}[tbp]
\centering
\includegraphics[width=\textwidth]{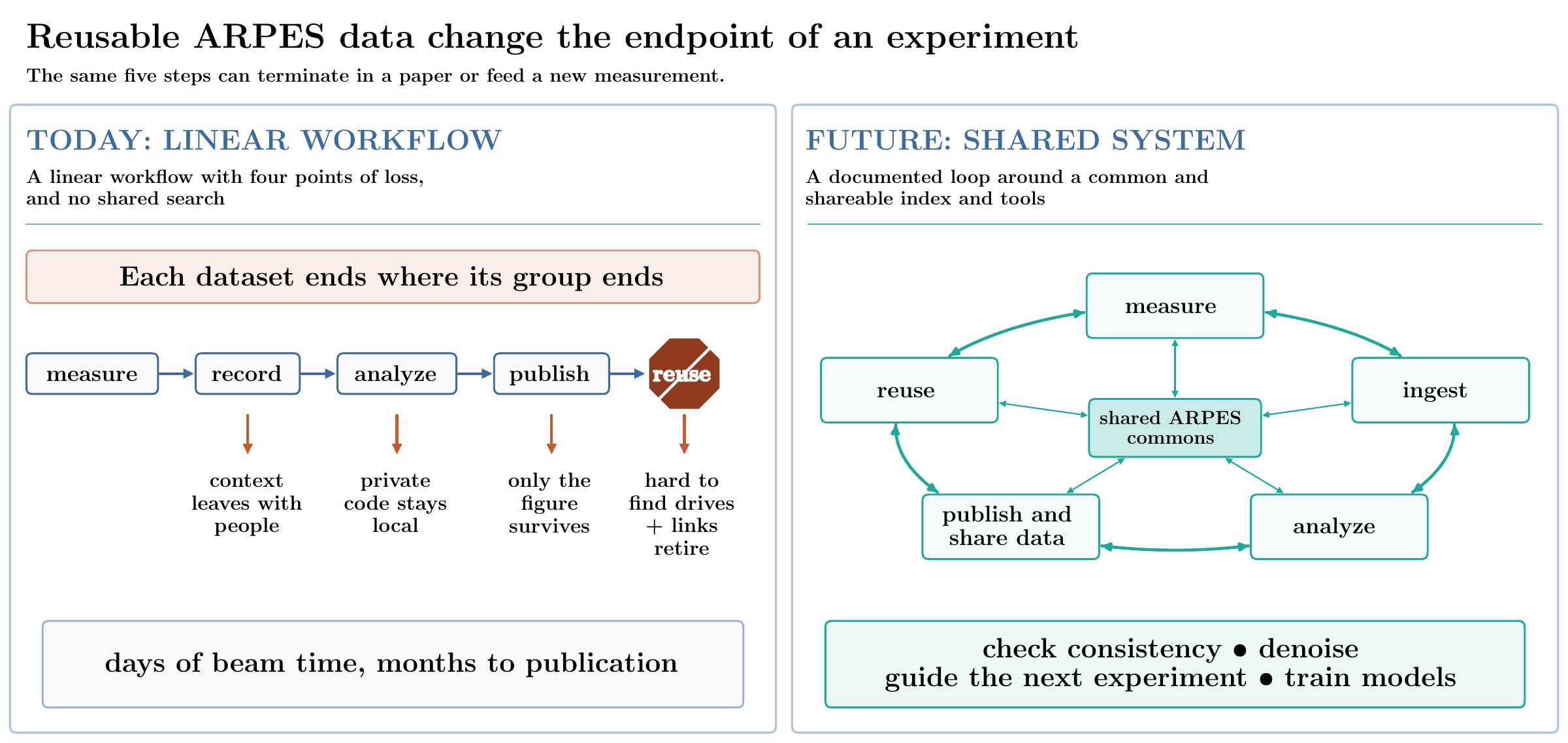}
\caption{The endpoint of an ARPES experiment, today and with a shared archive. Left: the prevailing linear workflow — measure, record locally, analyze with private code, publish a figure. Downward arrows mark where information leaks at each stage. After four leaks, each dataset ends where its group ends, and reuse is blocked. Right: the same five steps closed into a loop around a shared commons — standardized ingest, validated analysis tools, DOI-based sharing with notification tiers, and reuse by both people and models, so that any measurement can be checked for consistency, guide the choice of the next experiment, and train models. The change is organizational rather than instrumental: a common, searchable index with provenance, not new hardware.}
\label{fig:loop}
\end{figure*}

Because experimental context is currently discarded in disseminated ARPES data, there is a significant opportunity gap between the potential of ARPES and its data practice in reality (Fig.~\ref{fig:loop}). Today's workflow terminates largely in publication, while analysis code and much of the raw measurement data remain localized to specific laboratories, rather than in persistent scientific infrastructure. An experienced researcher knows, for example, what a bad cleave looks like, which weak feature is likely a matrix element effect, which anomaly has appeared before, which measurement condition unexpectedly worked, and which apparent discovery has already failed to reproduce across many other samples. When people leave the field, a substantial fraction of their informal, unstructured knowledge disappears. As in Fig.~\ref{fig:loop}(a), the conventional ARPES workflow is largely linear—measurement, record, analysis, publication—with information lost at each step. Published figures further select strongly for successful and interpretable measurements, while unsuccessful, ambiguous, and seemingly uninteresting spectra are rarely preserved, despite encoding valuable information about experimental conditions, failure modes, and the scientific target itself.

These losses will compound and become increasingly consequential as measurement throughput increases. The field faces a growing mismatch between what it can measure and measurements it is already capable of making. We intend to stem this loss of information and knowledge by creating a community-wide ARPES data \textit{ecosystem} \cite{oliveira2018}—an environment for creating, managing, sharing, and sustaining experimental data and their context—and encourage analogous efforts elsewhere in condensed matter physics. As illustrated schematically in Fig.~\ref{fig:loop}(b), the essential change is to make publication a waypoint rather than the endpoint of an experiment. A sufficiently rich database, ultimately including unsuccessful and ambiguous measurements alongside published results, would consolidate previously siloed and perishable experimental data into a shared, persistent framework. With both dataset volumes and publication activity still growing (Fig.~\ref{fig:landscape}), now is an opportune moment to establish a practitioner-driven standard before incompatible local conventions become entrenched.

\subsection{Immediate opportunities of the ecosystem — before foundation models}

Immediate benefits of the ecosystem flow straight to the community, independently of AI involvement. These include data preservation, findability, and credit. Published datasets themselves, not just the papers, become durable, searchable, citable research objects. Because a substantial fraction of historically important data resides on aging local storage, and in formats that depend on obsolete software whose data processing pipelines linking them to published results are not reproducible without extensive documentation, preservation would be vastly improved by standardization. These immediate benefits form a durable bedrock from which broader AI applications may be built.

Beyond preservation, the ecosystem could scale up existing workflows \cite{stansbury2020}. It would allow the community to learn from what papers currently leave out: ineffective, ambiguous, and seemingly ``uninteresting'' measurements provide negative examples and experimental knowledge that are rarely preserved in the published record, yet can be highly informative when viewed collectively. The ecosystem could function as a scientific time machine, allowing phenomena conceived or understood only in the future to be searched retrospectively across measurements taken years or even decades earlier (known elsewhere as \textit{precovery}) so that old experiments could yield genuinely new discoveries. A sufficiently broad corpus could act as an ``anomaly telescope'', directing attention to spectra that resist classification and may contain either hidden experimental problems or unexpected physics.

The most profound questions in ARPES involve recognizing patterns across material families, which depends on decades of accumulated personal experience. Beyond predefined material categories such as cuprates, kagome metals, or Weyl semimetals, important electronic behaviors and organizing principles may be distributed across chemically and structurally unrelated systems. An ecosystem could make that breadth of experience more accessible, allowing younger researchers to compare far more measurements and material classes than datasets available at any individual laboratory.

\section{From today's AI tools to modern AI-enabled science}

\subsection{What other data ecosystems teach us}
Structural techniques have had the most success so far in constructing community-wide databases, because of their compact representations and obvious downstream uses. Major examples include the Inorganic Crystal Structure Database (ICSD) \cite{ruhl2019} and PubChem \cite{kim2026}. Within condensed matter, non-structural databases include NIMS databases such as the Materials Data Repository \cite{mdr} for Hard X-ray Photoelectron Spectroscopy (HAXPES) and X-ray Absorption Fine Structure (XAFS), the X-ray Absorption Spectroscopy (XAS) Data Library \cite{xas}, and nmrXiv for Nuclear Magnetic Resonance (NMR) \cite{nmr}. However, the textbook precedent for building a community data ecosystem lies outside physics --- the Protein Data Bank (PDB), which collects and organizes protein sequences and structures, having grown over several decades into a database with over a quarter-million experimentally determined structures \citep{pdbhistory,wwpdb2026}. Recent breakthrough AI models such as AlphaFold were substantially trained on experimentally determined PDB structures, supplemented by self-distillation on predicted structures \citep{jumper2021}. \textit{We emphasize that the PDB was not created for AlphaFold,} but because structural biology needed a durable shared record. Transformative yet unanticipated downstream uses only emerged decades later.

ARPES differs from structural techniques and their databases (such as the PDB) in one fundamental respect: while structural results are reducible to a compact representation easily agreed upon by the community with well-developed validation conventions—encoding atomic or molecular structure—ARPES data lack an equally universal representation. As detailed in Table~\ref{tab:arpes}, spectral intensity is entangled with experimental context, and detailed analysis depends on bespoke analysis pipelines or models. An ARPES ecosystem must remedy this using standardized spectral data formats and benchmarks, with careful attention paid to tiering by data quality \cite{medallion}, while retaining original data in the database, as opposed to earlier structural databases in which only processed results were made available.

\subsection{Status of AI in ARPES today}
Despite increasing AI adoption in ARPES, the community currently relies on models operating within a pre-foundation model paradigm: they are trained on curated datasets for specific tasks with limited generalization to unseen, out-of-distribution data. Current applications broadly fall into three categories. (1) Analysis: AI-based denoising recovers useful structure from low-count spectra, including uncertainty estimates for quantitative analysis \citep{kim2021denoise,kim2026probdenoise}; simulation tools such as \textit{chinook} and synthetic ARPES pipelines support model development and training such as \textit{aurelia} \citep{day2019chinook,na2025simframework}; and automated methods are being developed for feature extraction, phase classification, and dispersion tracking \cite{peng2020,chen2026,xian2020,xian2023}. (2) Representation and generalization: representation learning and self-supervised embeddings enable few-shot classification and spatial-domain mapping with limited labels, while also exposing the need for a much larger and more diverse community dataset \cite{agustsson2025,Ekahana2023}. (3) Experiment control: autonomous micro-ARPES using Gaussian-process-guided search has already demonstrated AI-directed measurement in real and momentum space \citep{agustsson2024autonomous}, while broader efforts toward embodied experimental agents in quantum materials point to the next stage of ARPES automation \citep{melton2020,staab2024,qiu2026aims,shi2026qumus,staab2026}, towards closed-loop experiments where data collected in real time drives experiments forward without human involvement. We emphasize that these models would nevertheless already benefit from the construction of the ecosystem, by expanding the scale and diversity of their training data alone, even before foundation models are constructed.

\subsection{Towards future foundation models}
Robustness, the ability to generalize dependably out of distribution (that is, to cases not represented in the training set), depends critically on broader and more diverse training data, not only on model architecture. Models trained on narrow experimental or synthetic datasets struggle when tackling unfamiliar band structures, detector conditions, or sample states. As previous work such as \textit{aurelia} \cite{na2025simframework} have demonstrated, synthetic data can provide scale, but a large, diverse, and well-curated body of experimental data is essential for realism, validation, and reliable generalization. A foundation model approach offers a path beyond this limitation \cite{bommasani2022}. The model is first pre-trained on large, diverse datasets to learn general representations, and then adapted to specific tasks using smaller, high-quality experimental datasets. Because pre-training requires scale \cite{kaplan2020, hoffmann2022}, it may combine large synthetic corpora with increasingly broad collections of experimental spectra, while measured data remain essential for capturing the complexities omitted by simulation.

The more consequential possibility is not more robust performance on existing tasks, but qualitatively new capabilities that emerge from learning across a sufficiently broad corpus. Such capabilities could include recognizing phenomena unknown when the data were taken, identifying recurring patterns that define previously unnamed classes of quantum matter, and continuously ranking measurements by scientific anomaly. Further on, multimodal models could further connect ARPES with crystal structure from X-ray and neutron diffraction, scanning tunneling microscopy/spectroscopy (STM/STS), transport, inelastic scattering, optical spectroscopy, and theory, allowing relationships to emerge across measurements that are currently analyzed separately. In the longer term, linking such representations to crystal structures, synthesis data, and materials models could also enable inverse questions: not only what electronic structure a material possesses, but what materials might realize a desired electronic structure. Building a centralized, AI-ready global ARPES database aims to convert decades of fragmented experimental electronic structure data into a workable basis for foundation models for quantum materials and novel phenomena discovery. A summary of future possibilities in AI-enabled scientific discovery is given in Table~\ref{tab:foundation}.

Realizing this vision will require substantial community involvement to tackle questions \textit{en route} to foundation models. The gap between simulated and experimental spectra—including unknown self-energies, matrix elements, backgrounds, and instrumental artifacts—must be addressed during model development. There is an inherent trade-off: training on synthetic data promises scale, but propagates mistaken assumptions from the simulator to the data, while experimental data provide realism but remain noisy and limited. Unsupervised learning techniques, such as data augmentation and contrastive learning, as well as multimodal fusion to incorporate other experimental techniques, are promising avenues of research. The optimal architecture is also not yet known and should be designed around scientific data rather than inherited uncritically from language or computer vision. The central opportunity is therefore not a particular AI architecture, but a sufficiently rich and well-organized data foundation from which increasingly general scientific models can emerge.

\begin{table*}
\centering
\footnotesize
\caption{Possibilities in AI-enabled scientific discovery in condensed matter.}
\label{tab:foundation}
\renewcommand{\arraystretch}{1.25}
\begin{tabular}{p{4cm} p{12cm}}
\toprule
\textbf{Capability} & \textbf{Description} \\
\toprule
\raggedleft \textit{\textbf{Short}} & \textit{\textbf{Timescale}} \tabularnewline
\toprule

\raggedright Automated Spectral Analysis & Encode standard ARPES analysis protocols (MDC/EDC fitting, dispersion and gap extraction, self-energy determination) into reproducible pipelines that automatically extract the parameters experimentalists normally obtain by hand: tight-binding parameters, Fermi velocities, gaps, and self-energies, with calibrated uncertainties. \tabularnewline \midrule

\raggedright Systematic Theory Benchmarking & Use these extracted parameters to benchmark DFT, DMFT, GW, and quantum Monte Carlo against experiment systematically, scoring every spectrum in the database against competing theories with the same parameters and criteria rather than one case at a time. \tabularnewline \midrule

\raggedright  Cross Material Comparison & Uniform extraction across decades of data enables comparison of kink energies, gap structures, and band renormalizations across material families and control parameters (doping, strain, temperature, twist angle), distinguishing universal physics from material-specific features. \tabularnewline \midrule

\raggedright Anomaly Detection & Act as an ``anomaly telescope'' that continually ranks global experimental spectra to flag statistically extraordinary features, unexpected degeneracies, or unexplained spectral-weight transfers. Such tracking would allow the unanticipated discovery of new laws, thus new physics not envisioned in existing theory. \tabularnewline \midrule

\raggedright Collective Experimental Memory \& Open Access & Preserve tacit experimental wisdom (identifying bad cleaves, sample degradation, and instrumental artifacts) into persistent computational infrastructure, democratizing access to high-tier synchrotron or laser data for researchers globally. \tabularnewline

\toprule
\raggedleft \textit{\textbf{Long-Term}} & \textit{\textbf{Vision}} \tabularnewline
\toprule

\raggedright Closed-Loop Beamline Automation & Implement real-time and high speed/throughput active learning at synchrotrons and X-ray Free Electron Laser (XFEL) facilities to optimize measurement parameters dynamically, extract insights from ineffective/unpublishable data, and reanalyze historical datasets using modern calibration algorithms. \tabularnewline \midrule

\raggedright Multimodal Integration & Link ARPES data with complementary experimental techniques—such as STM, transport, optics, X-ray/neutron scattering to form a comprehensive model of empirical electronic structure model, which can be the basis for discovering unanticipated laws of physics not envisioned in theory; for example, any physical Big Data sets are expected to contain meta-laws such as multidimensional topological structures signaling novel higher-order emergent physics. \tabularnewline \midrule

\raggedright Universal Spectral Foundation Models & Pre-train AI on millions of spectra across diverse materials, photon energies, temperatures, and polarizations to capture generalized band structure representations while decoupling true quasiparticle dispersions from experimental noise and matrix-element and other technique-related artifacts. \tabularnewline \midrule

\raggedright Generative Inverse Materials Design & Invert the materials discovery pipeline by allowing scientists to specify a desired momentum-space feature (such as a topological flat band pinned at the Fermi level) while generative AI proposes the necessary chemical composition, space group, and substrate conditions.\tabularnewline

\bottomrule
\end{tabular}
\end{table*}

\section{Designing an ARPES data ecosystem for reuse and discovery}

Because important matters of debate should only be resolved in consultation with the broader ARPES community, the workshop also established an initial \textbf{steering committee} through an anonymous vote of participants. The committee will coordinate follow-up work, engage the broader community, and develop solutions to unresolved technical, policy, and governance questions. We summarize our discussion on metadata in Section IV\,A and carefully point out areas of convergence and disagreement in Section IV\,B. We anticipate that resolving technical and other differences would involve both the setting up of pilot databases (Section IV\,C) in consultation with the community guided by the steering committee.

\subsection{Organizing data and metadata}

A promising technical reference for organizing data is the NeXus/HDF5 framework \citep{nxmpesarpes}, and its application NXmpes for ARPES which builds on the FAIR data principles \cite{wilkonson2016}, in broad alignment with other segments of the condensed matter community \cite{scheffler2022}. This provides an existing community language for data and metadata which may be repurposed for the repository, rather than requiring the community to invent a new container from scratch. With the possibility of AI in mind, the community should consider a careful redesign of data infrastructure, while referencing existing efforts such as NEXUS/NXmpes, other existing technical foundations \cite{xian2020}, or other data management principles such as ALCOA and its successors \cite{rattan2017}. 

In addition, we propose dividing metadata into two layers. The first, a compact mandatory core, should be sufficient to interpret experimental geometry and reconstruct momentum coordinates (a derived quantity based on analyzer and beamline/light source setting). This would include essential information such as photon energy and polarization, measured angular coordinates and kinetic energy, and \textit{may} include other information such as analyzer and manipulator conventions, sample temperature, relevant resolutions, sample orientation, and schema version; specifics should be decided by the steering committee. The second layer comprises extensible context in the form of free text annotations, including (but not limited to) sample growth and handling, cleavage or preparation history, device state, pressure or field, detector calibrations, processing history, and experimental intent. Lastly, the community should adopt far more systematic means of metadata collection: all metadata for sample creation should be captured, such as end-to-end notebooks that spans sample preparers to ARPES measurements, and automation should be scaled up so that all metadata generated during measurement is captured by default in a standardized format. Attention should be paid to the capacity of data hosting platforms, in anticipation of a sharp growth in the volume of data collected.

\begin{table*}
\centering
\footnotesize
\caption{Principal issues and remedies identified in workshop discussion.}
\label{tab:issues}
\renewcommand{\arraystretch}{1.25}
\begin{tabular}{p{2cm} p{4.4cm} p{7.2cm} p{2.2cm}}
\toprule
\textbf{Issue} & \textbf{Why it matters} & \textbf{Possible remedy} & \textbf{Workshop status} \\
\toprule
\raggedleft \textit{\textbf{Data}} & \textit{\textbf{Deposition}} & & \\
\toprule

\raggedright Metadata burden &
A universal schema can become too complex and discourage participation, while
sparse metadata can make reuse impossible. &
Keep a small mandatory core (geometry/conditions); add optional extensions; version the schema; convert on ingest rather than forcing every instrument to change acquisition immediately. Retain free-text annotations and have more metadata fields (with the option to leave unfilled) to enforce standardization early. &
\raggedright Broad agreement \tabularnewline \midrule

\raggedright Raw versus processed data &
Raw data preserve future flexibility, but processed data may be easier to reuse
and much smaller. &
Use layered records: raw/native files to the extent feasible under storage capacity, standardized structured
data, and curated question-specific products. Keep provenance between layers. &
\raggedright Broad agreement, but open in detail \tabularnewline \toprule

\raggedleft \textit{\textbf{Data}} & \textit{\textbf{Analysis}} & & \\
\toprule

\raggedright Data quality is task-dependent &
Examples: spectra useful for band mapping may be inadequate for gap extraction or
Fermi-level determination; detector nonlinearities and surface aging can create
subtle artifacts; additional characterization of specimen/cleave, XPS, LEED, cleave photo, STM if in-situ; quality threshold of data to be posted &
Preserve calibration data and context; avoid a single database-imposed quality
score; attach task-specific quality flags and uncertainty; allow attributed
user comments; build automated checks that assist rather than gate deposition. The taking of calibration data in the first place may also be mandated. &
\raggedright Broad agreement \tabularnewline \midrule

\raggedright Simulation-to-experiment distribution shift &
Synthetic data are useful but inherit approximations in the generating theory
and instrument model. &
Flag simulated data explicitly; benchmark on measured held-out datasets; report
uncertainty and out-of-distribution performance; use real data as the anchor
for model validation. &
\raggedright Unresolved point \tabularnewline \midrule

\raggedright Data misuse &
Context-free spectra can be over-interpreted, or analyzed with inappropriate
corrections. &
Pair data with documented tools, provenance, owner notification or usage
statistics, and clear uncertainty or processing metadata and tools. Prefer education and
reproducible analysis over approval gates. &
\raggedright Broad agreement \tabularnewline \toprule

\raggedleft \textit{\textbf{Data}} & \textit{\textbf{Management}} & & \\
\toprule

\raggedright Credit \& attribution &
Depositors may lose visibility when data are reused at scale; sample growers and
facilities are easily under-credited. &
Persistent dataset identifiers; creator/contributor metadata including sample
source; required citation for identifiable dataset reuse; track usage. Develop a
separate credit mechanism for whole-corpus model training. &
\raggedright Crucial unresolved point \tabularnewline \midrule

\raggedright Access, ownership, \& facility policy &
Rights can depend on institutions, grants, facilities, proprietary agreements,
and embargoes. &
Support public, restricted, and non-distributable or training-only tiers; record
rights metadata; perform legal review by source type; do not assume the
depositor personally owns all data. Embargo period as an option. &
\raggedright Broad agreement: tiering supported \tabularnewline \midrule

\raggedright Hosting \& scale &
Actual storage volume and growth rate are not yet known; general repositories
have per-record limits (Zenodo defaults to 50~GB, with larger quotas possible)
\citep{zenodo}. &
Run a storage census before selecting infrastructure; pilot with representative
datasets; design for migration and mirroring; separate metadata/index services
from bulk-object storage. Question of distributed or centralized funding or hosting is unresolved. &
\raggedright Crucial unresolved point \tabularnewline \midrule

\raggedright Participation incentives &
Repository effort may fail if deposition demands extra work
while providing no local benefit. &
Make the standard useful behind laboratory firewall: fast retrieval,
automatic conversion, previews, shared analysis tools, one-click transition
from private staging to publication. A credit system may encourage good data upload by showing usage statistics, akin to GitHub. &
\raggedright Broad agreement, but open in detail \tabularnewline \midrule

\raggedright Governance \& maintenance &
A database can outlive the grant or institution that created it. &
Community governance, technical staff, stable funding,
international coordination, transparent change control, and periodic review of
standards and access policy. &
Broad agreement \tabularnewline
\bottomrule
\end{tabular}
\end{table*}

\subsection{Areas of convergence}

Table ~\ref{tab:issues} summarizes the principal issues raised at
the workshop, possible remedies, and the degree of
consensus reached. A primary topic of discussion was how data and metadata should be \textit{organized}. An option is to start with published data: these provide the lowest-friction pilot because their scientific results are already public and priority concerns are reduced. The matter of handling unpublished data warrants careful consideration. Additionally, data should be converted on ingest, so that no laboratory or beamline is required to abandon its established acquisition formats before the community resource becomes useful, and the onus of conversion should not fall on them. Whatever the source format, the archive should preserve raw measured coordinates and provenance rather than baking irreversible momentum-space conversions or processing choices into an only stored representation. Consistent with the goal of broad adoption, mandatory metadata core should be kept small while allowing extensions: the initial adopted standard should be practical enough to be adopted, rather than comprehensive enough to be complete. Free-text annotations should be treated as data in their own right, because experimental judgment and decision context are not fully captured by any fixed schema.

We also consider how the data should be hosted and used. Questions of credit and access are foremost concerns of the ARPES community, and should be treated carefully. Datasets should carry persistent identifiers and explicit credit, so that dataset citation becomes normal scholarly practice rather than lapse into being a mere informal courtesy. Access tiers should be supported, since pure openness cannot be the only path to participation when facility, incentive or institutional constraints differ. The community should first explore several pilot hosting sites before selecting permanent infrastructure. These pilots would provide empirical estimates of storage requirements, conversion effort, and metadata completeness. By having several independent pilot sites, early participation would be enhanced, and best practices may be identified for a future link up into an eventual integrated system. Lastly, benchmark tasks and evaluation criteria should be defined early, allowing different analysis and AI approaches to be compared systematically before broader claims about general-purpose foundation models are made.

\subsection{Scaling up the program}

Because many differences in technical implementation and community buy-in may can only be resolved by through small-scale working demonstrations, we propose a practical path in three phases, each with defined outputs and goals to hit before moving on to the next.

First, pilot databases are currently being set up by willing participants (see Supplementary Section~S9) to jump start the program and resolve teething problems early. These pilots should begin with published datasets contributed by multiple laboratories and facilities, spanning different instruments and local conventions. They should preserve native data, establish reusable converters into a common format, define a minimal metadata set sufficient to reconstruct geometry and provenance, and assign persistent identifiers, while avoiding premature attempts to impose a single experimental convention.

Next, in phase 2 the community should then use pilot databases to find and fix problems before they become ingrained in the larger ecosystem, testing whether the framework is sustainable, fair, and useful. In practice, this means estimating storage needs and securing a long-term hosting plan, agreeing on policies for attribution and access, and measuring the burden on contributors. This phase would include an end-to-end demonstration of the complete pipeline, from automatic metadata capture during an ARPES experiment to automatic upload into the database. Usefulness should be judged by whether researchers outside the contributing groups can find and reuse the data, and through a small set of benchmark tasks---such as cross-instrument dispersion extraction, uncertainty-aware denoising, or cross-material similarity search---with held-out experimental test sets and conventional analysis as a baseline.

Finally, in phase 3, the community would scale up once the framework is tried, tested, and trusted. Scaling would then proceed in three directions: extending automatic metadata capture from the demonstration site to other participating facilities; expanding from standard ARPES to specialty measurement modes; and adding private or pre-publication staging under the access policies agreed in phase 2, alongside a broader suite of community analysis and experiment support tools. As the corpus grows in diversity and quality, AI tools validated on the community benchmarks could feed back into data acquisition and adaptive measurement, and ultimately provide the basis for training future foundation models for ARPES.

\section{A practical path forward — and lessons beyond ARPES}

Many issues faced by the ARPES community—such as sample identity and condition, experimental conditions, and handling of metadata—are shared with the rest of condensed matter. Therefore, the implications of our proposal extend well beyond ARPES. Experimental condensed matter physics increasingly produces data that are richer than what can be conveyed in papers, while much of the knowledge required to interpret those data remains tacit and vulnerable to loss as researchers move on, instruments are decommissioned, and laboratories are shuttered. The very communities which understand these measurements should organize and preserve them now, while they can still determine what context matters and how it should be recorded. AI certainly promises new possibilities and lends urgency, but the case for building an ecosystem remains fundamentally scientific.

The data ecosystem would qualitatively transform the modus operandi by changing what constitutes the scientific record in condensed matter — and how science itself is carried out. Instead of disseminating knowledge through publications—whose results are selective interpretations of particular experiments—preserving reusable measurements together with the context needed to understand them would retain ambiguities, failures, alternative interpretations, and experimental conditions that would otherwise disappear because they were deemed unworthy of publication at time of collection. This would unlock new possibilities for data use. Measurements made with one scientific goal in mind may contain evidence relevant to problems conceived later, and patterns distributed across unrelated materials may reveal organizing principles that no individual experiment could expose. The ecosystem would turn the accumulated experience of a community into a durable collective memory, one that can be searched, revisited, recombined, and interrogated by questions that may not yet exist. We cannot know which future discoveries such an ecosystem will enable, for we can only connect the dots looking backwards. The most valuable questions the data may answer may very well be those we have not yet learned to ask.

\section*{Acknowledgements}

This research used resources of the Stanford Institute for Materials and Energy Sciences, supported by the DOE Office of Basic Energy Sciences, Division of Materials Sciences and Engineering under contract no. DE-AC02-76SF00515, the Stanford Synchrotron Radiation Lightsource, supported by the DOE Office of Basic Energy Sciences under contract no. DE-AC02-76SF00515, and Advanced Light Source, Supported by DOE Office of Basic Energy Sciences under contract no. DE-AC02-05CH11231. The Stanford Laboratory for Intelligent Matter provided funding for the workshop. We thank Sami Muhammad for helpful comments and editorial corrections.

%

\onecolumngrid
\clearpage
\newgeometry{textwidth=510pt,textheight=672pt,left=54.20251pt,top=52.27pt,headheight=25pt,headsep=16pt,footskip=30pt}
\makeatletter
\frontmatter@init
\let\maketitle\frontmatter@maketitle
\let@environment{abstract}{frontmatter@abstract}%
\patchcmd{\titleblock@produce}{\label{FirstPage}}{\label{SI-FirstPage}}{}{\typeout{SI: label patch not applied (harmless)}}
\patchcmd{\endthebibliography}{\label{LastBibItem}}{\label{SI-LastBibItem}}{}{\typeout{SI: label patch not applied (harmless)}}
\patchcmd{\NAT@bibsetnum}{\ref{LastBibItem}}{\ref{SI-LastBibItem}}{}{\typeout{SI: label patch not applied (harmless)}}
\makeatother
\setcounter{section}{0}\setcounter{subsection}{0}\setcounter{figure}{0}\setcounter{table}{0}\setcounter{equation}{0}\setcounter{footnote}{0}

\colorlet{RED}{red}
\renewcommand{\topfraction}{0.95}\renewcommand{\textfraction}{0.05}\renewcommand{\floatpagefraction}{0.75}

\renewcommand{\thesection}{S\arabic{section}}
\renewcommand{\thesubsection}{\Alph{subsection}}
\renewcommand{\thefigure}{S\arabic{figure}}
\renewcommand{\thetable}{S\arabic{table}}
\renewcommand{\theequation}{S\arabic{equation}}
\raggedbottom

\title{Supplementary Information for ``Framework for an AI-Ready ARPES Community Database''}

\date{October 10, 2026}

\begin{abstract}
This Supplementary Information documents the basis for recommendations presented in the main text and sets out how they could be implemented. Section~S1 describes the workshop, its source materials, and how community input is interpreted. Section~S2 details the methods behind the bibliometric survey of Fig.~2 in the main text. Section~S3 records current data practice in our groups and the lessons of earlier curation attempts. Section~S4 provides a compact technical account of why ARPES data require context (Table~I of the main text). Section~S5 proposes a provisional metadata specification and examines an example deposition. Section~S6 compares our data ecosystem with other data ecosystems. Section~S7 catalogs the AI applications presented in the main text, summarizes the lessons for foundation models, and proposes benchmark protocols. Section~S8 consolidates our discussion on access, credit, hosting, and governance into an operational policy table, and names the elected steering committee. Section~S9 lists goals, pilot deliverables, consensus statements, and unresolved decisions. Individual contributions to the discussion are described without attribution.
\end{abstract}

\maketitle

\section{Workshop methods and interpretation of community input}
\label{sec:workshop}

\subsection{Format, participants, and sources}

The inaugural Conference on Artificial Intelligence \& Electronic Structure was held at Stanford University on 4--6 August 2026 \cite{SI-aies2026}. Day~1 consisted of 23 invited presentations: two from domain experts in other fields (astrophysics and protein modeling), five from condensed matter theorists and non-ARPES experimentalists, and sixteen from ARPES experimentalists at various university laboratories and synchrotron user facilities in North America, Europe, and Asia. Every speaker was asked two questions in advance: (Q1) what their scientific goal was and whether an ARPES database would help, and (Q2) what would make them want to deposit their data. The collected answers are given in Secs.~\ref{sec:goals} and~\ref{sec:conditions}. Day~2 consisted of four structured discussion sessions (community end goals; contents, hosting, and governance; consensus building; action items) conducted on a shared online board, followed by a session to plan this paper.

Four sources are used to assemble this Supplement: Day~1 slides shared with the organizers (8 of the 23 talks); Day~1 recap and Day~2 refresh slides prepared by the organizers; export of the Day~2 board; and the main text. We note where a talk is referenced only through the recap or through notes taken during the talks. Table~\ref{tab:map} maps each section of this Supplement to the main text; we note that the steering committee elected at the workshop is listed in Sec.~\ref{sec:governance}.

\begin{table}[b]
\caption{Where each section of this Supplement supports the main text.}
\label{tab:map}
\begin{ruledtabular}
\begin{tabular}{p{1.4cm}p{6.9cm}}
Section & Main-text element supported \\
\hline
S1 & \cite{SI-aies2026}; Sec.~IV (steering committee) \\
S2 & Fig.~2 \\
S3 & Sec.~II\,B; Fig.~3 \\
S4 & Sec.~II\,A; Table~I \\
S5 & Sec.~IV\,A; Table~III (metadata, raw vs.\ processed) \\
S6 & Sec.~III\,A \\
S7 & Secs.~III\,B--C; Sec.~IV\,C (benchmarks); Table~II \\
S8 & Sec.~IV\,B; Table~III (credit, access, misuse, hosting, governance) \\
S9 & Secs.~IV\,C and V; Table~III (unresolved points) \\
\end{tabular}
\end{ruledtabular}
\end{table}

\subsection{How the Day~2 board is interpreted}
\label{sec:board}

\begin{figure}[ht]
\centering
\includegraphics[width=0.5\textwidth]{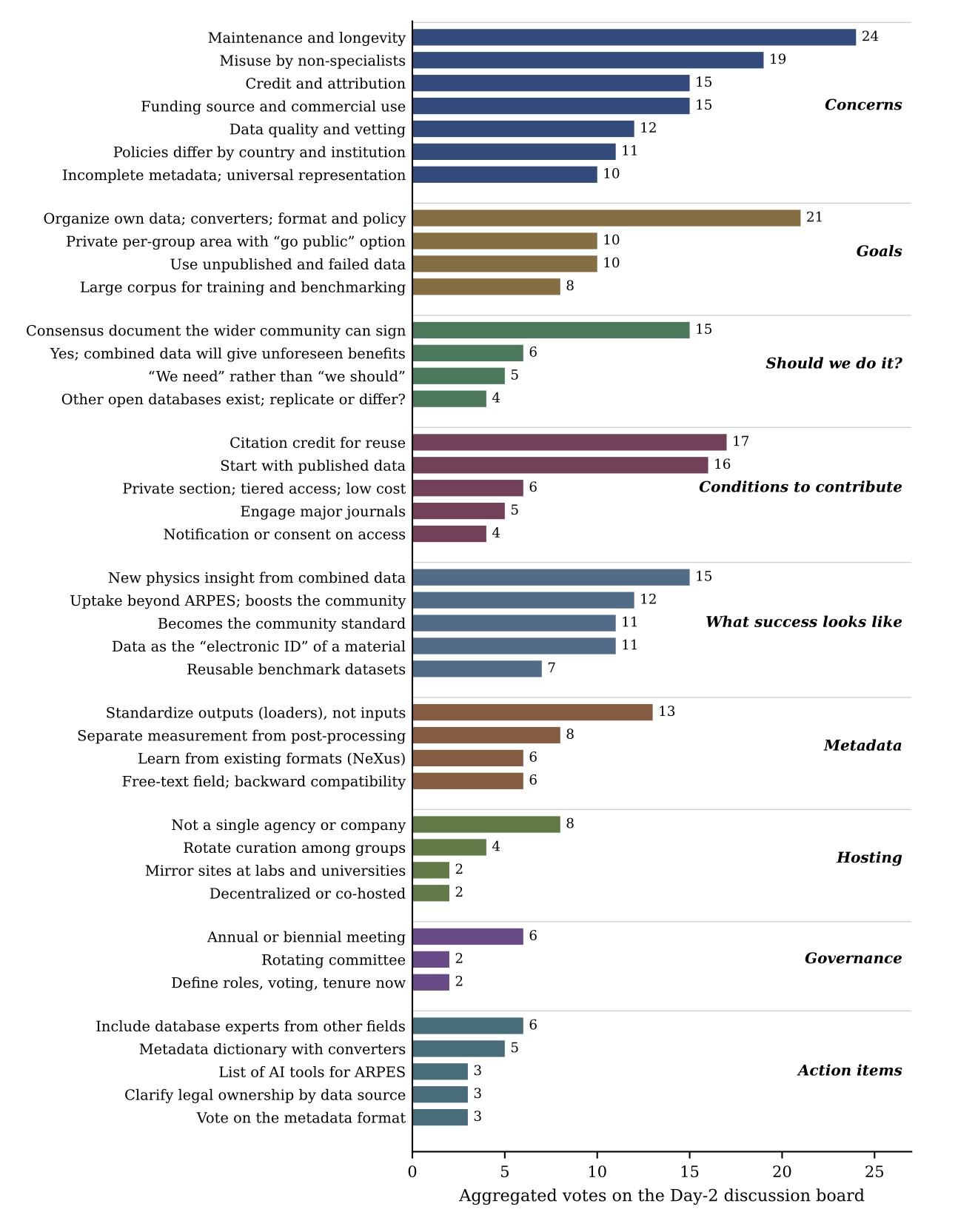}
\caption{Most-voted cards on the Day~2 discussion board, arranged by prompt. Counts are aggregated votes (Sec.~\ref{sec:board}); merged cards appear as single bars with their component statements listed in Secs.~\ref{sec:goals}, \ref{sec:metadata}, and \ref{sec:policy}. Logistical cards (meeting timing, communication channels) are omitted.}
\label{fig:votes}
\end{figure}

Day~2 discussions were structured around an online board on which comments would be placed on cards by participants and upvoted; the most upvoted comments became topics for immediate discussion. The board export, generated on 5~August~2026, records 187 cards, 127 comments, 617 votes, 50 participants, and 48 voters. Cards were anonymous and are not attributed here. We note three cautions regarding the vote counts quoted below. First, facilitators merged closely related cards, and summed their votes. A count on a merged card is therefore an aggregate of votes on its component statements (not a count of unique voters); we list the components where relevant. Second, votes were not exclusive: a participant could vote for several cards under one prompt. Third, the counts measure salience within a self-selected group of 48 voters. We therefore use the counts to indicate which issues mattered most to participants, not to establish consensus amongst the broader ARPES/condensed matter community. Figure~\ref{fig:votes} shows the most-voted cards under the substantive prompts.

Throughout this Supplement, we sort all recommendations into three categories: (i) a \emph{workshop-supported principle}: a position stated on the board or in talks that attracted substantial votes and no sustained objection; (ii) an \emph{participants' proposed implementation}: a concrete mechanism we propose here or in the main text as a way of realizing such a principle, which the workshop did not itself specify; and (iii) an \emph{unresolved option}: a question on which participants expressed incompatible positions or none. The ``Workshop status'' column of Table~III in the main text uses ``broad agreement'' in the sense of (i).

\section{Bibliometric methods for Fig.~2}
\label{sec:biblio}

Figure~2 of the main text reports 7{,}188 ARPES publications on quantum materials (1991--2025), 341{,}862 citations to them, and 726 distinct materials named in their titles, together with publication counts for four other experimental probes. This section records the procedure used to generate the numbers.

\emph{Source and date.} All counts were obtained from the Web of Science (Clarivate), current as of September~2026, with totals covering publication years 1991--2025.

\emph{Publication set, panel (a).} An ``All Databases'' topic search for ``angle-resolved photoemission'' and its spelling and acronym variants (ARPES; angle resolved photo emission; and related forms) was restricted by a keyword filter to quantum materials topics: correlated electrons and superconductivity, topology, quantum magnetism, and moir\'e and kagome lattices. Additionally, document types classed as preprints, Research Commons items, datasets, grants, and patents were excluded. The result contains 7{,}188 records.

\emph{Citations, panel (b).} Citations received by the records in (a) were binned by the calendar year in which they accrued rather than by the publication year of the cited paper, so that recent years are not suppressed by citation lag. The total count is 341{,}862.

\emph{Materials, panel (c).} Materials were identified by parsing the titles of records in (a). A material (system) is identified by its parent compound; distinct stoichiometries count separately (Bi$_2$Sr$_2$CuO$_6$ and Bi$_2$Sr$_2$CaCu$_2$O$_8$ are two materials); doping and strain collapse onto the parent (La$_{2-x}$Sr$_x$CuO$_4$ counts as La$_2$CuO$_4$ at every $x$); solid solutions resolve onto their independently studied end members (FeSe$_{1-x}$Te$_x$ counts as FeSe and FeTe); and a film on a substrate is distinct from the bulk (FeSe/SrTiO$_3$ and FeSe are two materials). Because only titles are parsed, a paper whose material is named only in the abstract is not counted, and the 726 distinct materials are a lower bound on the materials studied and an approximate count of those named.

\emph{Other probes, panel (d).} Publications per year were obtained for five probes under equivalent searches, each for the technique's name as an exact phrase without acronyms and restricted by the same quantum-materials filter; the resulting ARPES curve holds 85\% of the records in (a). ``X-ray spectroscopy'' covers x-ray absorption, emission, and dichroism; resonant inelastic X-ray scattering is counted under X-ray scattering; X-ray photoelectron spectroscopy is not included. X-ray and neutron search strings match families of related techniques whereas the ARPES string matches a single technique, so cross-technique ratios are only indicative.

\emph{Start year.} All panels begin in 1991, the first year for which the Web of Science Core Collection indexes abstracts and author keywords. A title-only control search showed that the large apparent jump in topic-search counts between 1990 and 1991 is an indexing artifact, thus earlier counts—which reflect title wording only—are omitted.

\section{Current data practice and lessons from earlier efforts}
\label{sec:status}

\subsection{What our groups reported}

Table~\ref{tab:practice} summarizes what participating groups reported about their own data holdings and practices on Day~1. The entries are self-reports from talks, cover only the groups that brought up this point, and are illustrative (rather than a comprehensive survey) --- the storage census called for in Table~III of the main text would extend this across the rest of the ARPES community. We note three main observations. \textit{First}, in one group which gave a full breakdown of their data holdings, the data volume that appeared in their papers was a small fraction of their archive, most of which consisted of alignment and spatial scans. Should this breakdown be common among the rest of the community, we anticipate that a shared archive should decide what to do with the remainder. \textit{Second}, even for recent data, the description of experimental geometry needed to reconstruct cuts in momentum space (Sec.~\ref{sec:context}) is often not recorded: one group reported that from its own archive only the analyzer cylinder geometry is reliably available, while physical dimensions, resolutions, cleave photographs, and (for old data) manipulator matrices are not. \textit{Lastly}, the facility-wide projections in the last row, of order one exabyte (EB) per year across all techniques at DOE light sources within five years, are not ARPES-specific and are included only to indicate the scale of the surrounding infrastructure.

\begin{table*}[!t]
\caption{Data practices reported by participating groups and facilities on Day~1. Self-reported and illustrative; groups are not identified. The last row is a facility-wide projection across all techniques, not an ARPES volume.}
\label{tab:practice}
\begin{ruledtabular}
\begin{tabular}{p{2.8cm}p{3.5cm}p{4.2cm}p{5.5cm}}
Group type & Volume reported & Storage and access & Notes \\
\hline
University lab, mixed ARPES and x-ray scattering & $\sim$ 7 TB total (2025); 50--500 GB per ARPES experiment & General-purpose repository (published); group drive or cloud folder (unpublished); one local PC and one RAID backup & Mixed txt/dat/fits and nxs/h5/krx formats; acquisition under LabVIEW, Python, vendor software, and SPEC/EPICS; transfer by SFTP or Globus; search via date, keyword, and logbook; metadata completeness x-ray diffraction $>$ scattering $>$ ARPES; institutional cloud too expensive \\
University lab, lab, synchrotron, and Tr-ARPES & $\sim$ 3 TB & Institutional repository for publication and thesis data since $\sim$2015 & Most volume consists of alignment and spatial scans; moving to reproducible data-to-paper notebooks; open analysis package with a typed metadata schema and analysis history \\
University lab, micro- and transport-ARPES & 5-D transport-ARPES set of $\sim$ 77{,}000 spectra; 22{,}800-point autonomous scan & Zenodo, unrestricted reuse, citation only & Training data for a published model are deposited but are not FAIR; metadata limited to the paper DOI \\
University lab, long-running synchrotron program & 26 years of archives & --- & Offered to contribute the archive; only analyzer cylinder geometry reliably recorded in old data \\
DOE Genesis Mission project & Corpus of 102 compounds; target 200+ & To be publicly released, with a non-redistributable tier & See Sec.~\ref{sec:spectra} \\
Light-source computing group & Projection of $\sim$ 1 EB per year across all DOE light-source techniques & Facility archives; facility ML platform & Beamline user data restricted to internal training unless joint projects with co-PIs \\
\end{tabular}
\end{ruledtabular}
\end{table*}

Mentions of other practices arose naturally and were brought up in the day's recap: group data on Zenodo (two groups), FAIR-structured group data, and facility-level collection efforts at a US light source. The recapconcluded: that data exist but nobody is yet depositing into a shared \emph{community} repository; that concerns about ownership, consent, credit, and incentives came up in nearly every talk; and that much data are in incompatible formats or in physical logbooks.

\subsection{Why earlier curation attempts did not persist}
\label{sec:lessons}

A Day~1 talk listed reasons that earlier attempts at ARPES data curation failed: dependence on what one participant idiomatically referred to as ``tender love and care'' to maintain their database; no incentive for the depositor to maintain their effort in the long run; an outsized ambition to capture every modality (strain, flux, and so on) from the outset; no quality control for soft metadata; and a data-collection process detached from archiving. The same talk listed why reuse of existing data in later publications rarely succeeds: a difficult search for the right kind of data, asking journals, asking authors ``upon reasonable request,'' checking DOIs, and asking for metadata. The talk's principle, ``make it dumb, then smarter'' (to standardize basic ARPES measurements and reduction protocols first and peel bespoke experiments off the initial effort) is the origin of the main text's insistence that the mandatory metadata core be small and that conversion happen on ingest, and of the sequencing in Sec.~IV\,C.

\section{Context dependence of ARPES data: a compact technical account}
\label{sec:context}

Table~I of the main text lists what must be preserved for each source of context dependence. Here we demonstrate this with the most common operations performed on ARPES data, the conversion of measured angles to momentum, and then distinguish measurement, processing, and interpretation.

\subsection{What momentum conversion requires}

An electron analyzer records counts as a function of kinetic energy $E_\textrm{K}$ and one or two emission angles. For a photoelectron leaving the surface at polar angle $\theta$ relative to the surface normal, the in-plane momentum is
\begin{equation}
k_\parallel = \frac{\sqrt{2 m_e E_\textrm{K}}}{\hbar}\,\sin\theta,
\label{eq:kpar}
\end{equation}
and the binding energy follows from $E_\textrm{B} = h\nu - \phi - E_\textrm{K}$ with work function $\phi$ \cite{SI-damascelli2003}. In Equation~(\ref{eq:kpar}), each symbol depends on recorded experimental context. $E_\textrm{K}$ must be referenced to the Fermi level, which requires either a reference spectrum (polycrystalline gold, or the sample itself) taken under the same analyzer settings, or the analyzer's energy calibration. $\theta$ depends on the analyzer's angular axis, the manipulator's polar, tilt, and azimuthal angles and their sign conventions, any angular offset determined from the data, and the orientation of the sample on the holder. Out-of-plane momentum depends on photon energy and an inner potential $V_0$, which is itself a fitted quantity. The intensity at each $(k,\,E_\textrm{B})$ is modulated by photon energy, polarization, and light-incidence geometry (matrix elements), energy and angular resolution, and detector response (including nonlinearity at high count rates). Finally, whether a spectrum is interpretable at all depends on surface quality and how it degrades over time---cleave quality, time since cleaving, temperature cycles---which are the information least often recorded.

In summary, lost metadata limits the usefulness of a spectrum. A dataset that lacks the manipulator conventions can still be preserved and searched, but it cannot be placed in momentum space; one that lacks a Fermi-level reference can yield dispersions but not gap sizes; one that lacks the resolution cannot be used to extract line widths. Main text therefore asks (Sec.~IV\,A) for a mandatory core sufficient to interpret experimental geometry and reconstruct momentum, and metadata specification in Sec.~\ref{sec:metadata} distinguishes what is required for admission to the archive from what is required for a particular scientific task.

\subsection{Measurement, processing, and interpretation}

Three layers should be kept distinct in any record. The \emph{measurement} is the array of electron counts on its native axes (analyzer energy, angles) together with the instrument state. \emph{Processing} comprises reversible or irreversible operations applied afterwards: Fermi level alignment, nonlinearity and transmission function corrections, normalization, angle-to-momentum conversion, symmetrization, and background subtraction. \emph{Interpretation} comprises fitted quantities (peak positions, line widths, gaps, velocities) and model-dependent inferences such as a self-energy relative to an assumed bare band. The Day~2 board proposed that measurement metadata should be recorded separately from any post-processing (8 votes), with processing history stored inside the data file or alongside it. The board also agreed that data quality is task dependent (for example, a spectrum may be ``good for band mapping but not gap analysis''), that the archive should not impose a single quality score, and that data ``bad'' for one purpose carry information useful for another (a model trained to find a good spot on a sample needs realistic bad data). One card proposed ``published'' data as the one definitive marker, with subjective tags reserved for user comments and ratings. These positions underlie the ``Raw versus processed data'' and ``Data quality is task-dependent'' rows of Table~III; we note that the processing level of a record (raw, standardized, curated) is a statement about what has been done to it, not about its scientific reliability.

\section{Provisional metadata specification and an example deposition}
\label{sec:metadata}

The main text proposes a compact mandatory core sufficient to reconstruct geometry and momentum, with extensible free text annotations, and leaves the specific field list to the steering committee. Here we give a provisional specification (Table~\ref{tab:metadata}) assembled from Table~I of the main text, the mandatory core list in Sec.~IV\,A, and the Day~2 discussion, so that the pilot has a concrete starting point. We emphasize that this is the participants' proposed implementation in the sense of Sec.~\ref{sec:board} rather than a community standard --- we expect the steering committee to modify it in consultation with the rest of the ARPES community.

\subsection{Three metadata tiers}

Although the main text presents a two-tier distinction of metadata, the Day~2 discussion deviated from this. The most-voted metadata card (13 votes) argued against standardizing the \emph{input} side, because facilities' formats have inertia, and for standardizing the \emph{output} side through community-supported loaders into existing open analysis platforms; a comment proposed to gather representative files from all groups, report the current level of uniformity, and generate converters with group-specific recommendations, which another comment called ``the only reasonable proposal which would not create too high of a barrier.'' A second card (2 votes) proposed an agreed mandatory baseline plus an architecture for optional and future structured fields (for example, sample characterization, device parameters such as gate voltage, post-processing such as nonlinearity correction) with future dimensions added by the governing body. A third (6 votes) argued for a free-text field from the outset, because ``we do not yet know what metadata will ultimately be important'' --- with a long comment arguing that a fixed schema shaped by early use cases can bias later searches, that free-text notes should remain as important as structured fields, and that contextualization created during reuse should be contributed back as attributed, revisable layers. Taken together these suggest three tiers: a compact \emph{structured core}, \emph{structured extensions} that are optional but typed when present (field, pressure, device state, calibrations), and \emph{free-text context}. The second layer of the main text combines the latter two.

Two further conventions were proposed on the board and are adopted in Table~\ref{tab:metadata}: a versioned schema, so that the conventions in force for a time-stamped dataset are unambiguous, with each group publishing its own notation alongside its data; and explicit ``unknown,'' ``estimated,'' and ``not applicable'' values, so that legacy data with incomplete metadata can be admitted without pretending to a completeness they lack. Two cards (2 votes each) stressed that the effort of adding metadata to old data ``cannot be too high.''

\begin{table*}[!t]
\caption{Provisional metadata specification for the pilot, organized by field group. ``Core'' fields are required for admission to the archive; ``extension'' fields are optional but structured when present; ``context'' is free text. Values may be recorded as known, unknown, estimated, or not applicable. The right-hand column indicates which fields a representative scientific task additionally requires, to make explicit that admission and task-readiness are different thresholds. The specification is the authors' proposal and is to be mapped onto the NeXus NXmpes\_arpes application definition~\cite{SI-NXmpesarpes} during the pilot; a \texttt{.nxs} file extension by itself does not establish a common meaning for its contents.}
\label{tab:metadata}
\begin{ruledtabular}
\begin{tabular}{p{1.6cm}p{4.1cm}p{4.1cm}p{2cm}p{3.5cm}}
Field group & Core (required for admission) & Structured extension (when available) & Free-text context & Required additionally for the task of\ldots \\
\hline
Identity and provenance & Dataset identifier; sample identifier; contributors (including who measured and who grew the sample); instrument or beamline; acquisition timestamp; schema version; conventions document & Cleave or surface identifier linking measurements on the same surface; parent multidimensional dataset; related datasets; publication DOI & Experimental intent; logbook excerpt & Credit (sample source); leakage control in benchmarks (cleave and session identifiers) \\
Signal and coordinates & Intensity array on native axes; axis names and units; for each axis whether measured or derived; energy scale type (kinetic or binding) & Fermi-level reference spectrum or calibration; derived momentum axes with the parameters used; detector-image coordinates & --- & Gap extraction (Fermi reference); momentum-space comparison (derived axes with parameters) \\
Experimental conditions & Photon energy; polarization; sample temperature; manipulator angles and sign conventions; analyzer angular axis and light-incidence geometry & Energy and angular resolution; analyzer lens mode, pass energy, slit; spot size; pressure; sample bias & Alignment procedure & Momentum reconstruction (full geometry); linewidths and self-energies (resolutions) \\
Sample and surface & Composition or parent compound; crystallographic orientation if known & Growth and preparation; cleave method and time since cleave; measurement position on sample; aging or spectral-quality evolution; reference XPS, LEED, or cleave photograph & Cleave quality remarks; anomalies noticed & Cross-sample comparison (preparation, time since cleave) \\
Sample environment & Flag for any non-equilibrium condition & Applied field, strain, current, optical excitation, gate voltage, with the instrument used to set and read each & --- & Any environment-dependent study \\
Processing & Processing status (raw, standardized, curated) and a link to the less-processed record & Each operation with software, version, and parameters: calibration, nonlinearity and transmission corrections, normalization, interpolation, symmetrization, background & Analysis narrative & Reproduction of a published figure (full history) \\
Rights and release & Access tier; rights holder or data-source type; permitted reuse & Embargo date; dataset version and change log; usage statistics & --- & Training-only use (permission to train recorded separately from permission to redistribute) \\
\end{tabular}
\end{ruledtabular}
\end{table*}

\subsection{Existing tools}

The Day~1 recap described the main existing ARPES database---the NXmpes NeXus application definition as the standard, pynxtools as the conversion pipeline, and NOMAD as the hosting platform, with public datasets already hosted---and noted that adoption by a broad audience has been slow. The NeXus documentation distinguishes the general NXmpes definition for photoelectron spectroscopy from the ARPES-specific NXmpes\_arpes definition, which extends it and requires a defined ARPES coordinate system, the analyzer geometry, a transformation chain for sample orientation, and a data block of one energy and two angular axes; the main text's ref.~\cite{SI-NXmpesarpes} cites the latter. Table~\ref{tab:metadata} is designed to be expressible in that definition. Two of the participants described in-house tooling that already implements parts of it: an open-source notebook-driven analysis package with a typed metadata schema and a recorded analysis history, and a converter from several instruments and beamlines to a single NeXus-based file together with an acquisition workflow in which standard operations ($E$--$k$ cuts, Fermi-surface maps, domain and signal-to-noise optimization) are exposed as agent-callable tools and an automated check validates metadata before upload. A funding program that would support one to two staff to travel among groups building converters was also described. Under action items, a metadata dictionary with converters received 5 votes, with comments pointing to pynxtools and NeXus as the starting point.

\subsection{A deposition walk-through}
\label{sec:walkthrough}

Phase~1 deliverables of Sec.~IV.C in the main text (preserve native data; reusable converters; a minimal metadata set sufficient for geometry and provenance; persistent identifiers) translate into a series of actions to take to upload a representative dataset of standard (equilibrium) ARPES data. Of these actions, a community data loader can perform a subset automatically. Ingesting the native files and any notes unchanged, converting signal, axes, and whatever instrument metadata the native format carries into the schema (with absent fields set to ``unknown''), validating units and the presence of core fields, and assigning a DOI, contributor records, and an access tier are all automatic once a loader exists for the source format. The other two steps cannot be done by the depositor alone. Supplying the core fields that the native format lacks---geometry conventions, the Fermi reference, temperature, sample identity---requires the depositor, who should mark values recovered from notes or memory as estimated; for legacy data this is the main cost, and it is why two cards insisted that the effort for old data ``cannot be too high.'' Reproducing a published figure from the standardized record, where the dataset underlies one, requires the depositor's processing history and an expert comparison in which discrepancies are recorded rather than silently corrected. Later corrections create a new version with a change log rather than overwriting, so that the DOI remains citable. The validation report and the log of manual interventions from these steps are the data that the ``independent reuse'' benchmark of Sec.~\ref{sec:benchmarks} needs.

\section{Lessons from other data ecosystems}
\label{sec:ecosystems}

The main text (Sec.~III\,A) argues that structural techniques built community databases first because their results reduce to a compact, agreed representation, and that ARPES lacks one. Table~\ref{tab:ecosystems} collects the precedents raised at the workshop, the lesson each offers, and where the analogy to ARPES breaks down. We comment on two precedents in more detail.

The history of the Protein Data Bank~\cite{SI-BermanBurley2025,SI-Barinaga1989} was reviewed on Day~1: deposition was voluntary from 1971, became a condition of publication only after a petition by senior structural biologists in the 1980s, and before that\textcolor{red}{,} scientists deposited because the data would never be lost, because they wanted errors caught, and for the public good; the archive survived four changes of dominant technique and is itself biased toward what crystallography measures easily. The most-voted card under \emph{Should we do it?} (15 votes)---a document recording the consensus of those present that the wider community can later sign---led to the consensus decision to write this paper. NexusLIMS~\cite{SI-Taillon2021}, a laboratory information management system for shared-use electron microscopy, is closer to ARPES practice than the PDB: a co-op rather than a user facility, with little enforceable top-down control, vendor-specific formats, and datasets abandoned over time; its design lesson was to build modularly on institutionally supported infrastructure and existing practices so as to survive changes underneath.

\begin{table*}[!t]
\caption{Data ecosystems raised at the workshop, the lesson each offers for ARPES, and the limitation of the analogy.}
\label{tab:ecosystems}
\begin{ruledtabular}
\begin{tabular}{p{2.9cm}p{6.4cm}p{6.4cm}}
Ecosystem & Transferable lesson & Limitation of the analogy \\
\hline
Protein Data Bank & Voluntary deposition can precede mandates if depositors gain preservation, error checking, and credit; a format that can evolve (Diamond $\to$ PDB $\to$ mmCIF) lets the archive outlive techniques; transformative uses (AlphaFold) arrived decades later and were not planned for & A structure is a compact, validated result; an ARPES spectrum is a context-dependent measurement with no agreed reduced form, so the archive must retain original data, which the PDB did not need to \\
Inorganic crystallography (ICSD) & Deposition attached to publication as an easy step, with deep integration into analysis software; structure determination has a convention for what is deposited & Same as above; also, condensed-matter spectroscopy has no deposition culture outside crystal and band structure \\
Neutron scattering & Natively centralized collection and analysis at facilities, with standardized reduction protocols for ``basic'' measurements; community analysis codes that grew from a standardized core & ARPES is split between user facilities and home laboratories, with soft metadata (surface, cleave, charging) that scattering largely lacks \\
Materials Project, NOMAD, arXiv & API access for programmatic reuse; longevity as a public good; demonstration of productivity and primacy through timestamps & Computed data are regenerable and uniform; experimental spectra are neither; arXiv curates documents, not multidimensional records \\
Light-source knowledge graph~\cite{SI-FAIR2WISE} and medallion tiers~\cite{SI-medallion} & Link materials, publications, data, and metadata so that a record is ``queried with a specific scientific question in mind''; refine data in tiers (bronze, silver, gold) with provenance between them & A processing tier is not a quality grade: more processed does not mean more reliable, and task-specific quality must be recorded separately (Sec.~\ref{sec:context}) \\
NexusLIMS~\cite{SI-Taillon2021} & A co-op facility with no enforceable top-down control can still curate if the system is modular and built on existing institutional infrastructure and practices & Electron microscopy data are images with vendor metadata; ARPES adds geometry and surface context that no vendor format carries completely \\
Zenodo and institutional repositories & Already used by several of our groups; DOIs and citation-only reuse work today & Per-record limits (50 GB default, larger on request \cite{SI-zenodo}); no ARPES-aware validation, search, or loaders; rate-limiting of bulk download \\
\end{tabular}
\end{ruledtabular}
\end{table*}

The Day~2 board asked how astrophysics handles misuse of public data by non-specialists, and one action item (6 votes) proposed including people from biology or astrophysics with database experience to audit the pilot, with computer scientists experienced in large training corpora also suggested; the consensus prompt proposed inviting both groups to the next meeting. Notes from the Day~1 astrophysics talk record four uses of AI there---acquisition, emulating expensive processes, learning posteriors, and shared representations---with calibrated uncertainty and full instrument fidelity as requirements, and a framing of theory-driven forward models versus data-driven foundation models as a bias-versus-variance trade-off.

\section{AI use cases, benchmark design, and validation limits}
\label{sec:ai}

\subsection{What was presented}

Table~\ref{tab:apps} lists the AI applications presented or proposed at the workshop, their evidence status, and what each needs from a shared archive. The last column is our inference from the presentations and the board, not a statement by the presenters, and is intended to show that existing tools would benefit from the archive even before the existence of foundation models (main text Sec.~III\,B, last sentence). One card on the board noted that many published ARPES tools did not release code, and an action item (3 votes) proposed compiling a list of AI tools for ARPES together with the standard computational and statistical tools that an agent would call as back-ends.

\begin{table*}[!t]
\caption{AI applications in ARPES presented or proposed at the workshop. Evidence status: \emph{published} (peer-reviewed or preprint result shown), \emph{presented} (unpublished result or prototype shown in a talk), \emph{proposed} (described as a plan or raised on the board). The last column is the authors' inference.}
\label{tab:apps}
\begin{ruledtabular}
\begin{tabular}{p{4.0cm}p{1.9cm}p{4.4cm}p{5.2cm}}
Application & Status & What was shown & What it needs from the archive \\
\hline
Denoising of low-count spectra, including uncertainty estimates & Published & Recap only \citep{SI-kim2021denoise,SI-kim2026probdenoise} & Paired low- and high-count measurements with resolution metadata; held-out materials \\
Correction of detector nonlinearity & Presented & Recap only & Retained calibration data per instrument and session \\
Feature extraction, dispersion tracking, phase classification & Published & Recap only \cite{SI-peng2020,SI-chen2026,SI-xian2020,SI-xian2023} & Standardized axes; task-specific quality flags \\
Convolutional autoencoder representation of spectra~\cite{SI-Agustsson2025} & Published & $256\times256$ spectra compressed to a 512-dimensional latent; compression, denoising, clustering; representation preserved across materials; training data on Zenodo without metadata & Spectra from many materials on a common image grid; material and condition labels to test what the latent encodes \\
Self-supervised latent space for x-ray scattering~\cite{SI-Choudhary2026}, proposed for ARPES & Published (scattering); proposed (ARPES) & 83.6-million-parameter conditional VAE on 1.5 million images; one latent space serves denoising, stopping prediction, anomaly flagging at ingest, similarity search & A large unlabeled corpus; ingest-time hooks so that quality flags are computed on deposit \\
Autonomous micro-ARPES~\cite{SI-Agustsson2024} & Published & Gaussian-process-guided search over position and angle; 22{,}800-point example & Spatial maps with positions retained; records of the acquisition decisions \\
Beam-spot quality scoring; sample-edge finding; acquisition loop with simulator and on-the-fly analysis & Presented & Recap and notes only & Realistic ``bad'' data labeled as such; intent metadata distinguishing optimization scans from measurements \\
Agentic laboratory workflow with vision-language model and metadata validation & Presented & Converter to NeXus; standard operations exposed as agent tools; automated metadata check before upload & A schema to validate against (Table~\ref{tab:metadata}) \\
Domain-adversarial bridge from simulated to measured spectra~\cite{SI-He2021} & Published & ``Train with theory, deploy on experiment'' & Simulated records flagged as such; matched experimental records \\
Fast spectral simulator (10 ms per spectrum) & Presented & Prototype; initial CNN tests; related effort of \cite{SI-na2025simframework} & Experimental data to measure the domain gap against \\
Self-energy prediction from structure and DFT, trained on ARPES & Proposed, in progress & See Sec.~\ref{sec:spectra} & Raw spectra over wide $k$ ranges with full conditions; extracted self-energies where available \\
Search for topological invariants; cross-modal ``quantization-proximity'' search & Proposed & Program combining spin- and time-resolved ARPES, STM/STS, and nonlinear optics & Links from ARPES records to STM and optics records on the same material \\
Similarity search, next-experiment suggestion, automatic characterization (``electronic forest'') & Proposed & Each material carries its CIF and measured Fermi surface; a generative model ``trained on the forest, not on papers'' & Material identity and crystal structure attached to every record; momentum-space products \\
Standard AI agent on the acquisition computer & Proposed (board, 6 votes) & Context-dependent notes during beamtime, capturing decisions & A free-text context layer that tools can write to \\
\end{tabular}
\end{ruledtabular}
\end{table*}

\subsection{Lessons from other foundation models}
\label{sec:foundation}

A supplementary lecture on dataset construction drew lessons from computer vision, language modeling, and biochemistry. It argued that the bottleneck for foundation models in condensed matter ``is the assembly of large datasets across experimental techniques --- which is a political problem first, and only a technical problem second.'' We summarize four lessons relevant to the main text.

\emph{Distribution shift is not overfitting.} Replication studies in image classification~\cite{SI-Recht2019,SI-Taori2020} showed that a new test set built by the same procedure as the original produces an accuracy drop that is a distribution shift distinct from overfitting, that models keep their relative ranking under such shifts while humans are unaffected, and that robustness to synthetic perturbations does not transfer to natural shifts. The questions this raises for ARPES are: which shifts to expect (between instruments, between measured-easily and measured-rarely systems, between techniques) and whether multimodal or contrastive training across techniques mitigates them. This is the basis of the main text's statement that robustness depends on breadth and quality of data.

\emph{Data quality is operational.} In the DataComp benchmarks~\cite{SI-Gadre2023,SI-Li2024}, ``quality'' is defined by downstream performance rather than by human judgment, which tends to overvalue polished data and undervalue diversity; the ``bitter lesson''~\cite{SI-Sutton2019} makes the same point for hand-built priors. For ARPES this means an instinctive judgment of ``good data'' should not gate deposition, which \sout{is what} the board concluded independently (Sec.~\ref{sec:context}).

\emph{Synthetic data are useful but limited.} Synthetic data can supply scale when real data are scarce, but they propagate the simulator's assumptions and, when models train on their own outputs, risk model collapse~\cite{SI-Kazdan2025,SI-Geng2024}. The lecture's own suggestion was to pre-train on spectra from DFT, DMFT, or tight-binding models across as many systems as possible and fine-tune on measured data; the self-energy project of Sec.~\ref{sec:spectra} follows the same sequence. What remains open is the \emph{proportion and fidelity} of synthetic data: what fraction of pre-training is synthetic, how faithful the simulator must be to matrix elements and instrument response, whether simulated spectra enter the archive itself (one card proposed adding them, explicitly labeled, especially for systems beyond DFT), and how much experimental supervision is needed before predictions on unmeasured materials can be are all open topics. Table~III of the main text records the agreement that simulated data be flagged and that measured data anchor validation.

\emph{Data attribution could inform credit, but is not yet a credit system.} Methods that estimate the contribution of individual training examples to a model's output---Data Shapley~\cite{SI-Ghorbani2019}, influence functions~\cite{SI-Koh2017}, datamodels~\cite{SI-Ilyas2022,SI-Ilyas2025}---were presented as a way to quantify the value of specific datasets so that ``researchers' hard-won data do not just disappear into a black box,'' and to locate gaps where data should be taken. These methods are computationally demanding, model-specific, and unvalidated as measures of scholarly contribution; we regard them as an exploratory addition to be used alongside the traceable dataset manifests and contributor records proposed in Sec.~\ref{sec:policy}.

\subsection{A worked example: self-energy prediction trained on ARPES}
\label{sec:spectra}

One presentation described a DOE Genesis Mission Phase~I project combining ARPES, oxide molecular-beam epitaxy, correlated-electron theory with machine learning, and micro-ARPES at a national-laboratory beamline. It is the most concrete existing plan to use community ARPES data as training data.

The goal is to predict the self-energy $\Sigma(k,\omega)$ from crystal structure and DFT band structure alone, using ARPES spectra in supervised learning. The corpus consists of raw or published spectra placed on a common $(k,\,E)$ grid together with CIF and DFT entries; it held 102 compounds at the time of the workshop, with a target of 200 or more, 10 held out for testing. This number is the project's assembled corpus of compounds with training-usable data; it should not be confused with the 726 distinct materials named in ARPES publication titles since 1991 (main text Fig.~2), most of which exist only as figures, and the presentation's point was that ``most published ARPES lives inside figures, and vision models are bad at them.'' The model is a small residual U-Net, pre-trained on simulated spectra and then fitted to the measured corpus, with physical constraints imposed: a non-negative quasiparticle scattering rate (with the retarded convention, $\Gamma = -\operatorname{Im}\Sigma^{R} \geq 0$), Kramers--Kronig consistency between real and imaginary parts, and crystal symmetry. The planned test is to grow a compositional-gradient wafer of a correlated material never measured by ARPES (e.g., La$_{1-x}$Sr$_x$VO$_3$ across the Mott transition) and to measure hundreds of compositions by micro-ARPES with a spot below 10~$\mu$m, comparing with prediction.

In this example project, it is important that \textit{what quantities measured, fitted, and inferred} should be kept distinct. The archive itself holds measured spectral intensities, while dispersions and line widths are fitted from momentum or energy distribution curves, and the project noted that the model itself must decide which analysis applies. The self-energy is then inferred relative to a bare band that is never measured and is taken from DFT, an approximation with known failure modes in correlated systems; the inferred $\Sigma$ is further affected by matrix elements, resolution, overlapping bands, $k_z$ broadening, and the limited energy window of a given measurement, and the project's initial assumption of a local (momentum-independent) $\Sigma$ ``will be bad for certain materials'' such as the cuprates and the orbital-selective iron chalcogenides. Different groups may extract different self-energies for the same material, which is part of what the model must absorb.

Two levels of participation were offered: public release, in which processed data are used for training and then published in the released corpus; and non-redistributable, in which data are used for training but not released, although measurement metadata are. The trained model is to be released in either case. This separates permission to train from permission to redistribute, which is the distinction adopted in Table~\ref{tab:policy}.

\subsection{Benchmark protocols and the leakage problem}
\label{sec:benchmarks}

The main text names three benchmark tasks for the pilot---cross-instrument dispersion extraction, uncertainty-aware denoising, and cross-material similarity search---with held-out experimental test sets and conventional analysis as the baseline, and adds that usefulness should be judged by whether outside groups can find and reuse the data. Table~\ref{tab:bench} shows the authors' proposals on how each would be evaluated.

It is important to remember that spectral cuts from the same sample, the same cleave, the same measurement session, or the same parent multidimensional dataset are not independent, and a model that has seen one will perform unrealistically well on the others. Held-out sets must therefore be defined at the level of material, sample, or instrument, not spectrum; the cleave and parent-dataset identifiers in Table~\ref{tab:metadata} exist to make this possible. Spectrum counts therefore overstate the number of independent materials, therefore the corpus sizes in Sec.~\ref{sec:spectra} should be read with that in mind.

\begin{table*}[!t]
\caption{Proposed evaluation of the benchmark tasks named in Sec.~IV\,C of the main text. Authors' proposals.}
\label{tab:bench}
\begin{ruledtabular}
\begin{tabular}{p{3.2cm}p{6.2cm}p{6.2cm}}
Task & Evaluation & Safeguard \\
\hline
Cross-instrument dispersion extraction & Error in band position and Fermi velocity against conventional MDC/EDC fitting on the same data; fraction of cases in which the reported uncertainty interval covers the conventional value & Hold out entire instruments or laboratories, not spectra; report performance separately for in-distribution and held-out instruments \\
Uncertainty-aware denoising & Preservation of gap magnitudes, line widths, and weak features against an independent higher-count measurement of the same surface; calibration of reported uncertainties & Test for introduced features and for systematic distortion of line shapes; never evaluate against a denoised reference \\
Cross-material similarity search & Retrieval of spectra of the same material or the same spectral feature across changes of photon energy, temperature, and instrument & Test whether similarity tracks physics or instrument identity by querying with the same material measured on different instruments \\
Independent reuse & Time and assistance an outside group needs to locate a dataset and reproduce a published dispersion or figure from it & Log every missing metadata and manual intervention; these logs feed back into the core-field list \\
Automatic metadata capture (Phase~2) & Fraction of core fields populated without manual entry during an ordinary beamtime at the demonstration site & Compare with the fraction populated for legacy data of the same instrument \\
\end{tabular}
\end{ruledtabular}
\end{table*}

\section{Access, attribution, hosting, and governance}
\label{sec:policy}

This section summarizes the Day-2 discussion on how the archive would be used and run. It supports Sec.~IV\,B of the main text and the ``Data misuse,'' ``Credit and attribution,'' ``Access, ownership, and facility policy,'' ``Hosting and scale,'' ``Participation incentives,'' and ``Governance and maintenance'' rows of Table~III. The policy choices are collected in Table~\ref{tab:policy}; the discussion that led to them follows.

\begin{table*}[!t]
\caption{Operational policy choices the pilot must test, with the options raised on the Day-2 board. Status follows Sec.~\ref{sec:board}: WP, workshop-supported principle; AI, authors' proposed implementation; UO, unresolved option.}
\label{tab:policy}
\begin{ruledtabular}
\begin{tabular}{p{2.4cm}p{6.2cm}p{4.8cm}p{2cm}}
Choice & Options raised & What the pilot should test & Status \\
\hline
Public download & Anonymous download; registration required (``a minimal gateway to stop bot scraping''); per-access notification of the depositor; per-access permission & Whether registration plus notification satisfies depositors without deterring reuse & WP (open with registration); UO (per-access permission) \\
Restricted access & Project-specific collaboration with clear rules; joint projects with co-PIs for facility user data; invitation-only registration at first & Whether a restricted tier attracts data that would otherwise not be deposited & WP (tiers) \\
Training-only use & Data used to train a released model but not redistributed, with metadata released (Sec.~\ref{sec:spectra}) & Separate permissions for (a) training, (b) redistributing the data, (c) releasing model weights; whether depositors use the tier & AI \\
Embargo & Private per-group staging with a ``go public'' control (10 votes); a maximum embargo of 3 or 5 years so that data are not forgotten; delay to defuse competition & Whether staging raises deposition of unpublished data; the embargo length & WP (staging); UO (length) \\
Correction and versioning & Versioned schema; corrected datasets as new versions with change logs; who may correct & Whether versioning preserves citability & AI \\
Contributor credit & Dataset DOI; required citation for identifiable reuse (17 votes); creator records naming who measured and who grew the sample; acknowledgment in the main text of papers using the data; facilities credited for infrastructure, not interpretation & Whether citation of datasets appears in practice; whether student and postdoc contributors are findable & WP \\
Corpus-level credit & Citation of the database as a whole; manifests listing every dataset used to train a released model; usage statistics shown to depositors; data attribution as an exploratory measure (Sec.~\ref{sec:foundation}) & Whether a manifest and usage statistics are regarded by depositors as adequate credit for whole-corpus training & UO \\
Commercial and industry use & Whether commercial entities may train on the data; whether industry funding would entitle their models to the data (15 votes, no resolution) & To be expressed through the rights metadata of Table~\ref{tab:metadata} once a position is reached & UO \\
Legal ownership & Rights depend on institution, grant, facility, and country (11 votes); review by data-source type; ``should we involve an actual lawyer'' & A per-source-type rights review before scale-up & WP \\
\end{tabular}
\end{ruledtabular}
\end{table*}

\subsection{Conditions for contributing}
\label{sec:conditions}

Day~1 answers to Q2 (\emph{What would make you want to deposit your data?}) were:
\begin{itemize}
    \item An explicit access, reuse, and attribution policy, with guidelines for poor legacy metadata,
    \item Settle ownership, consent, and credit before pooling; lower the threshold and add incentives (reuse projects, funding, publication mandates),
    \item Governed, curated data with a semantic layer of intent and provenance,
    \item Limited, project-specific collaboration rather than open-ended pooling,
    \item Standardized labels for material, doping, and temperature,
    \item Facility beamline data for internal training only, with sharing beyond that through joint projects,
    \item FAIR HDF5 at acquisition, with published or unused data merged into a repository,
    \item ``Unrestricted deposit already works for me: no permission needed, just cite the paper'',
    \item And ``the database simply needs to exist; model quality is capped by dataset diversity.''
\end{itemize}

On Day~2, the most-voted conditions were: \begin{itemize}
    \item Citation credit for reuse (17 votes, merged from four cards: citation for reuse in publication with whole-corpus training excepted,
    \item Credit to the students and postdocs who took the data,
    \item Inclusion of sample growers,
    \item Explicit acknowledgment in the main text of papers) and starting with published data (16 votes: ``let's start with something simple that gets traction'',
    \item ``Phase 1, published data; Phase 2, unpublished later'').
\end{itemize}

A private section with tiered access received 6 votes. Talking directly to the major journals to encourage or require deposition received 5 votes but drew objections (forcing commitment at the start is hard to maintain, and much good ARPES is published elsewhere) and we recorded it as unresolved. Per-access consent (4 votes) was the most-commented card and the comments were mostly opposed: posting carries the understanding that someone may download; whole-corpus AI cannot work under per-access consent, so ``if we hope to generate new knowledge in this way, we have to be quite open''; gatekeeping could be a depositor option, ``but if the majority adopt this protocol, the database loses its purpose''; notification without permission would still be valued. Cards under other prompts added that a converter from familiar text files ``would greatly encourage adoption'' (4 votes) and that the pilot should ``quickly demonstrate, even with a simple example, what people can actually do with the database'' (3 votes).

One Day~1 talk gave the most specific incentive list: recognize and fund projects based on data reuse; conference sessions, fellowships, and data-mining hackathons; counting a group's data effort toward beamtime proposal ratings; and, as enforcement, publication requirements. It also listed what the effort needs (data, computing, storage, full-time staff) and what a depositor might receive (investment, beamtime).

\subsection{Credit}

The credit card under \emph{Concerns} (15 votes, 14 comments) produced the longest thread on the board. The distinction that emerged (and Table~III of the main text adopts) is between credit for identifiable reuse of a few datasets in a paper (for which a DOI and a required citation suffice), and credit for whole-corpus training, ``where individual accesses are much less the main point.'' Mechanisms proposed for the former are: an ARPES-specific Zenodo-like service; a DOI per dataset; a shared credit policy modeled on large high-energy-physics collaborations; a depositor-selected reuse license. Questions left open were: who is credited (experimenters, PI, growers); what credit means for a student (a citation or a CV line); who is ``first'' (to measure, upload, publish, or conceive); and how deposited data become citable in scholarly indices. On primacy, one comment proposed that the deposit timestamp establishes who measured first, while priority for an interpretation continues to rest on publication or preprint. One comment cited a pending lawsuit by a news publisher over training use of its articles as a caution.

\subsection{Misuse and quality}

The second most-voted concern (19 votes) was misuse by non-specialists: ``strangers use data in an unauthorized way, producing questionable science.'' Remedies proposed in comments ranged from education (a visualizer that applies the common corrections, bridging data with documented analysis tools, peer review as the existing gatekeeper) to gated access (registered users only, mandatory discussion with the dataset creator, owner notification). The observation that already-published data are already public narrowed the concern to unpublished data. Table~III records the preference for education and reproducible analysis over approval gates as broad agreement; the gating proposals were individual positions. Data quality and vetting (12 votes, 10 comments) is treated in Sec.~\ref{sec:context}. Lower-voted concerns included: who can claim to be first (3); whether success with language models will ``put ARPES experimentalists out of business'' (2); conflicting data from different groups and inconsistency between runs (2 each); the activation barrier of using another group's data under different norms (1); biases in dataset weighting (1); and how to define the boundary of the ARPES community, given non-specialist beamtime users and adjacent techniques.

\subsection{Hosting}
\label{sec:hosting}

The most-voted concern overall (24 votes, 15 comments) merged maintenance, management, and a comment on ``moderation and oversight, so that [the ecosystem] doesn't die and become obsolete in a couple years.'' Under \emph{Who should host?}, no positive option received strong support; the most-voted card (8 votes) was a negative preference, ``not owned/hosted by a single federal agency/company,'' and one comment warned that national-laboratory hosting ``may be dangerous given the current situation.'' Options raised included: pooled resources to hire professionals versus rotation among groups (rotation received 4 votes but comments noted unequal manpower and weak enforceability); mirror sites at laboratories and universities (2 votes); decentralized or co-hosted across institutions (2); a few pilot sites first; existing services such as Zenodo (``not Google'') subject to per-record limits and rate-limiting of bulk downloads; national computing and data resources (DOE model-training consortia, the American Science Cloud, the American Data Cloud), whose suitability was not evaluated. Two comments observed that database efforts are already under way in some countries, so that a single archive may be impossible for policy reasons and a superstructure connecting parallel efforts may be unavoidable. One card asked that storage need and cost be defined before a hosting model is chosen, with a census in which each group reports what it would submit today; the survey questions proposed for it are in Sec.~\ref{sec:actions}. The main text's recommendations---a storage census, several independent pilot sites, and designing for migration and mirroring---follow from this.

\subsection{Governance}
\label{sec:governance}

Under \emph{Who should be responsible?}, the most-voted card (6 votes) was a yearly or biennial meeting like the present one; others proposed a rotating committee, a committee of four to six whose groups also maintain the database during a two-to-three-year tenure, inclusion of students and postdocs ``who really get their hands dirty with the AI work,'' and a request that roles, responsibilities, voting, and tenure be defined. Responsibilities suggested: access-control and quality-control policy (``representing the community for decisions, not doing all the work''; 3 votes); exploring funding for hosting; organizing annual meetings. Structural suggestions with votes: draw from both national laboratories and universities, with both theorists and experimentalists, and a chair or co-leads with experience of both ARPES and databases (5); include the computing and AI-for-science side (4); co-lead by two or three people combining experiment, data, and AI (4); an international advisory committee alongside a domestic executive committee (4). One card argued for defining roles and drafting the document before selecting people, so that the broader community can weigh in.

Nominations were collected through a form and on the board; we do not reproduce the nominations or their vote counts. An initial steering committee was established by anonymous vote of participants before the workshop closed. Its members are Tanny Chavez (Advanced Light Source), Andrea Damascelli (University of British Columbia), Yu He (Yale University), Chunjing Jia (University of Florida), Takeshi Kondo (University of Tokyo), Laurenz Rettig (Fritz Haber Institute), Eli Rotenberg (Advanced Light Source), Kyle M. Shen (Cornell University), Zhi-Xun Shen (Stanford University), and Inna Vishik (University of California, Davis). The committee's remit is that stated in Sec.~IV of the main text: to coordinate follow-up work, engage the broader community, and develop solutions to the unresolved technical, policy, and governance questions listed in Sec.~\ref{sec:openpoints}. Its term, rotation, and eventual confirmation by the wider community were not settled at the workshop.

\section{Pilot deliverables, consensus statements, and unresolved decisions}
\label{sec:open}

\subsection{Goals and definitions of success}
\label{sec:goals}

Day~1 answers to Q1 (\emph{What is your goal?}) were:
\begin{itemize}
    \item Seed an ARPES foundation model by self-supervised pre-training on curated facility data,
    \item A self-driving laboratory that understands ARPES and picks the next measurement,
    \item A foundation model for quantum matter that predicts regimes nobody has measured,
    \item Robust denoising of low-count trARPES, which needs diverse, realistic training data,
    \item Causal-discovery AI linking growth, spectra, and properties,
    \item Rescue three decades of beamline data and restart AI-friendly with a facility upgrade,
    \item Preserve group data and enable real reuse, with FAIR metadata and provenance from data to paper; repeat biology's path by building the curated corpus first,
    \item A theorist-accessible, analysis-ready database fed automatically by the beamline,
    \item Extract Hamiltonian parameters directly from spectra, then automate it with agents; automate ``basic'' ARPES so that a standardized archive builds itself during measurement,
    \item Compact learned representations for 5-D, terabyte-scale datasets,
    \item And an inference approach to the trARPES energy--time resolution trade-off.
\end{itemize}The recap grouped these as (1) a transferable foundation model, (2) closing the acquisition loop, and (3) new insight from pooled past data, which correspond to the long-, intermediate-, and near-term timescales of the main text. Several of these are aspirations whose assumptions and validation remain to be specified; the benchmark protocols of Sec.~\ref{sec:benchmarks} are a first step towards this.

On the board, the most-voted goal (21 votes, merged from four cards) was to organize one's own group's data in a FAIR-compatible manner within months regardless of the workshop outcome, build a converter to a standard searchable format, build both a unified dataset and a generalizable prescription for experimental datasets across condensed matter, and at minimum decide a format, a location, and a policy to replace journals' ad hoc upload rules. Making use of unpublished and ``failed'' data received 10 votes, and a large corpus for training and benchmarking 8. Definitions of success, in order of votes: a physics insight from the combined data that human researchers had overlooked (15 votes); uptake beyond ARPES specialists, with the effort boosting rather than killing the community (12); the database becoming the standard for all ARPES groups (11); experimental data becoming the ``electronic ID'' of a material, compared systematically with DFT, DMFT, and QSGW (11); reusable benchmark datasets analogous to the standard test images of computer vision (7); shared, adaptable acquisition and analysis scripts (4); validation or refutation of conclusions from other data during peer review, and a database so easy to use that no group can plausibly claim deposition is unfeasible (3); and a quantifiable success metric (3). One Day~1 talk called for a ``poster child'' demonstration---an ``AlphaFold moment''---and the Day-2 refresh posed the same question.

\subsection{Action items and the status-quo survey}
\label{sec:actions}

Action items with corresponding vote counts are listed here:
\begin{itemize}
    \item Include database-experienced people from biology or astrophysics to audit the process (6 votes),
    \item A metadata dictionary with converters (5),
    \item A list of AI tools for ARPES (3),
    \item Clarify legal ownership by data source (3),
    \item A democratic vote on the metadata format (3),
    \item Survey all groups' data files and compare fields (2),
    \item Each group publishes its conventions with its data (2).
\end{itemize}
Without votes: estimate each group's upload volume and server cost; define quantifiable success metrics and explicit incentives; investigate national data-cloud resources; a small committee with defined roles and term limits; a literature search on what other fields have done; start with old published data. Among next steps, setting up a test server and uploading the easiest files first received 7 votes, with comments to check existing repositories first and to review the literature before fixing a schema; a proposal for a digital-twin framework of instrument geometry and $k$-space coordinates was shelved for later.

Questions proposed for a community status-quo survey are: each group's ARPES data volume in TB; the split of time between user-facility and laboratory ARPES; a ranking of the biggest challenges (infrastructure, incentive, technological); and representative data files, contributed by link, for a longitudinal comparison of metadata fields. These questions are the template for the storage census in Table~III of the main text, and their answers would extend Table~\ref{tab:practice} across the community.

\subsection{Scope and consensus statements}

A dedicated prompt asked whether the database should hold published data or any ARPES data with good metadata: the latter received 5 votes and the former 1, while a card proposing separate purpose-specific databases received none and three objections (hard to maintain; one database with intent flags has better longevity; the separation is itself a bias). Together with the 16-vote ``start with published data'' card (Sec.~\ref{sec:conditions}), this suggests that published data come first as a \emph{sequencing} decision, with the eventual scope being any adequately described data in a single archive with intent flags.

Cards under \emph{Consensus} were: another meeting next year (12 votes), in August; at least one paper describing the findings and plans (3); a first committee to take the next step (2); and invitations to computer scientists and to PDB or AlphaFold people. A six-point plan (2 votes)---work toward a database; lead by the working group but bring the broader community together; a white paper as the first step to identify stakeholders; a few database sites, connected or separate, usable for benchmarking; a community-approved steering committee with terms and governance; and eventually funding, a sustainable platform, and policy---closely tracks the program laid out in Secs.~IV\,C and V of the main text.

\subsection{Unresolved decisions}
\label{sec:openpoints}

Here we list the questions that the workshop left open, and that the steering committee and the pilot are to resolve, are collected here; they correspond to the ``unresolved'' and ``open in detail'' entries in Table~III of the main text.
\begin{enumerate}
\item The content of the mandatory metadata core, the level of details required, and whether imaging and time-of-flight analyzers share a schema (Sec.~\ref{sec:metadata}).
\item Per-access permission versus open download with registration and notification (Sec.~\ref{sec:conditions}).
\item Whether to seek journal deposition mandates now or let the archive earn adoption (Sec.~\ref{sec:conditions}).
\item Credit for whole-corpus model training, beyond manifests and usage statistics (Secs.~\ref{sec:foundation} and \ref{sec:policy}).
\item Commercial and industry access (Table~\ref{tab:policy}).
\item The proportion and fidelity of synthetic data in pre-training, and whether simulated spectra enter the archive (Sec.~\ref{sec:foundation}).
\item The permanent hosting model and host, after the storage census and pilots (Sec.~\ref{sec:hosting}).
\item The steering committee's term, rotation, and confirmation by the wider community (Sec.~\ref{sec:governance}).
\item The boundary of the ``ARPES community'' for deposit, access, and voting, given non-specialist beamtime users and adjacent techniques (Sec.~\ref{sec:policy}).
\end{enumerate}

%

\end{document}